\documentclass[twocolumn]{aastex701}

\usepackage [english]{babel}
\usepackage [autostyle, english = american]{csquotes}
\MakeOuterQuote{"}

\usepackage{amsmath}

\usepackage{graphicx}
\usepackage{xspace}

\graphicspath{{./}{figures/}}
\makeatletter
\def\input@path{{tables/}{figures/}}
\makeatother

\newcommand{\TESS}{TESS\xspace}

\newcommand{\Exofast}{\texttt{EXOFASTv2}\space}

\newcommand{\Rstar}{\ensuremath{R_{\star}}\xspace} 

\newcommand{\Rjup}{\ensuremath{R_\mathrm{J}}\xspace} 
\newcommand{\Mjup}{\ensuremath{M_\mathrm{J}}\xspace}
\newcommand{\Mnep}{\ensuremath{M_\mathrm{Nep}}\xspace}
\newcommand{\Rearth}{\ensuremath{R_\oplus}\xspace} 
\newcommand{\Mearth}{\ensuremath{M_\oplus}\xspace}
\newcommand{\Rsun}{\ensuremath{R_\odot}\xspace} 

\newcommand{\Rp}{\ensuremath{R_p}\xspace}
\newcommand{\Mp}{\ensuremath{M_p}\xspace}
\newcommand{\rhop}{\ensuremath{\rho_p}\xspace}
\newcommand{\Teff}{\ensuremath{T_\mathrm{eff}}\xspace}
\newcommand{\logg}{\ensuremath{\log{g}}\xspace}
\newcommand{\feh}{\ensuremath{\mathrm{[Fe/H]}}\xspace}
\newcommand{\vsini}{\ensuremath{v\sin{i}}\xspace}

\newcommand{\ms}{\ensuremath{\mathrm{m}\,\mathrm{s}^{-1}}\xspace}

\newcommand{\gcc}{\ensuremath{\mathrm{g}\,\mathrm{cm}^{-3}}\xspace}

\begin{document}

\title{Discovery of an Inflated Hot Neptune and Its Formation from Jovian Mass Loss
\footnote{This paper includes data gathered with the 6.5 meter Magellan Telescopes located at Las Campanas Observatory, Chile}}

\correspondingauthor{Grant~C.~Weldon}
\author[0000-0003-4081-1839]{Grant~C.~Weldon}
\email{gweldon@astro.ucla.edu}
\affiliation{Department of Physics \& Astronomy, University of California Los Angeles, Los Angeles, CA 90095, USA}
\affiliation{Mani L. Bhaumik Institute for Theoretical Physics, Department of Physics and Astronomy, UCLA, Los Angeles, CA 90095, USA}

\author[0000-0001-7961-3907]{Samuel~W.~Yee}
\email{syee@astro.ucla.edu}
\altaffiliation{51 Pegasi b Fellow}
\affiliation{Department of Physics \& Astronomy, University of California Los Angeles, Los Angeles, CA 90095, USA}

\author[0000-0001-7840-3502]{Bradley~M.~S.~Hansen}
\email{hansen@astro.ucla.edu}
\affiliation{Department of Physics \& Astronomy, University of California Los Angeles, Los Angeles, CA 90095, USA}

\author[0000-0002-9802-9279]{Smadar~Naoz}
\email{snaoz@astro.ucla.edu}
\affiliation{Department of Physics \& Astronomy, University of California Los Angeles, Los Angeles, CA 90095, USA}
\affiliation{Mani L. Bhaumik Institute for Theoretical Physics, Department of Physics and Astronomy, UCLA, Los Angeles, CA 90095, USA}

%% Ran initial blend analysis
\author[0000-0001-8732-6166]{Joel D.\ Hartman}
\email{jhartman@astro.princeton.edu}
\affiliation{Department of Astrophysical Sciences, Princeton University, 4 Ivy Lane, Princeton, NJ 08544, USA}

%% Part of original proposals
\author[0000-0002-4265-047X]{Joshua N.\ Winn}
\email{jnwinn@princeton.edu}
\affiliation{Department of Astrophysical Sciences, Princeton University, 4 Ivy Lane, Princeton, NJ 08544, USA}

\author[0000-0003-1305-3761]{R.~Paul~Butler}			% PFS Team
\affiliation{Earth and Planets Laboratory, Carnegie Institution for Science, 5241 Broad Branch Road, NW, Washington, DC 20015, USA}
\email{bluaper@gmail.com}

\author[0000-0002-5226-787X]{Jeffrey~D.~Crane}			% PFS Team
\affiliation{The Observatories of the Carnegie Institution for Science, 813 Santa Barbara Street, Pasadena, CA 91101, USA}
\email{crane@carnegiescience.edu}

\author[0000-0002-5674-2404]{Phil Evans}				% SG1: El Sauce
\affiliation{El Sauce Observatory, Coquimbo, 1870000, Chile}
\email{phil@astrofizz.com}

%% Ground-based photometry
\author[0000-0002-4503-9705]{Tianjun~Gan}
\affiliation{Instituto de Astrof\'{i}sica de Canarias (IAC), E-38205 La Laguna, Tenerife, Spain}
\affiliation{Departamento de Astrof\'isica, Universidad de La Laguna (ULL), E-38206 La Laguna, Tenerife, Spain}
\email{tianjungan@gmail.com}

\author[0000-0002-2532-2853]{Steve~B.~Howell}			% SG3: Gemini/Zorro
\affiliation{NASA Ames Research Center, Moffett Field, CA 94035, USA}
\email{steve.b.howell@nasa.gov}

\author[0000-0001-9269-8060]{Michelle Kunimoto}
\affiliation{Department of Physics and Astronomy, University of British Columbia, 6224 Agricultural Road, Vancouver, BC V6T 1Z1, Canada}
\email{mkuni@phas.ubc.ca}

\author{David~Osip}										% PFS Team
\affiliation{Las Campanas Observatory, Carnegie Institution for Science, Colina el Pino, Casilla 601, La Serena, Chile}
\email{dosip@carnegiescience.edu}

\author[0000-0003-2196-6675]{David Rapetti}
\affiliation{NASA Ames Research Center, Moffett Field, CA 94035, USA}
\affiliation{Research Institute for Advanced Computer Science, Universities Space Research Association, Washington, DC 20024, USA}
\email{david.rapetti@nasa.gov}

\author[0000-0002-8681-6136]{Stephen~A.~Shectman}		% PFS Team
\affiliation{The Observatories of the Carnegie Institution for Science, 813 Santa Barbara Street, Pasadena, CA 91101, USA}
\email{shec@carnegiescience.edu}

\author[0000-0002-3481-9052]{Keivan~G.~Stassun}
\affiliation{Department of Physics and Astronomy, Vanderbilt University, Nashville, TN 37235, USA}
\email{keivan.stassun@vanderbilt.edu}

\author[0009-0008-2801-5040]{Johanna~K.~Teske}			% PFS Team
\affiliation{Earth and Planets Laboratory, Carnegie Institution for Science, 5241 Broad Branch Road, NW, Washington, DC 20015, USA}
\affiliation{The Observatories of the Carnegie Institution for Science, 813 Santa Barbara Street, Pasadena, CA 91101, USA}
\email{jteske@carnegiescience.edu}

\author{Roberto~Zambelli}  								% SG1: LCO - 2195
\affiliation{American Association of Variable Star Observers, 185 Alewife Brook Parkway, Suite 410, Cambridge, MA 02138, USA}
\affiliation{Societ\`{a} Astronomica Lunae, Castelnuovo Magra, Italy}
\email{robertozambelli.rz@libero.it}

%% CHIRON data reduction
\author[0000-0002-4891-3517]{George~Zhou}
\affiliation{University of Southern Queensland, Centre for Astrophysics, West Street, Toowoomba, QLD 4350 Australia}
\email{George.Zhou@unisq.edu.au}

%% High angular resolution imaging
\author[0000-0002-0619-7639]{Carl~Ziegler}				% SG3: SOAR
\affiliation{Department of Physics, Engineering and Astronomy, Stephen F. Austin State University, 1936 North Street, Nacogdoches, TX 75962, USA}
\email{Carl.Ziegler@sfasu.edu}

\begin{abstract}
The production of Neptune-like planets with orbital periods of 3--6 days is challenging for conventional models of high-eccentricity migration. We present the discovery and characterization of TOI-2195~A~b, an inflated hot Neptune ($P = 4.16$ days, $m_p= 1.46M_{\rm Nep},\,R_p = 0.79R_{\rm J}$) orbiting an early K-type star with a wide binary companion at $\sim 600$~au. Detection of the Rossiter-McLaughlin effect at $\sim2.6\sigma$ confidence with Magellan/PFS reveals the planet is likely on a near-polar orbit with a sky-projected stellar obliquity $\lambda = {109^{+35}_{-53}} ^{\circ}$. We perform coupled dynamical and structural modeling that reproduces the observed characteristics of the system. We show that the planet may have originated as a cold, Jovian planet that was excited to high eccentricities via the stellar Eccentric Kozai-Lidov (EKL) mechanism, where it lost up to $\sim90\%$ of its mass via Roche lobe overflow during close periastron passages, enabling rapid tidal migration and radius inflation due to tidal heating. TOI-2195 A b provides a test for planetary migration theories, and our simulations suggest that puffy hot Neptunes originated as more massive Jovians that underwent mass loss during high-eccentricity migration.

\end{abstract}


\section{Introduction} \label{sec:intro}

The formation of sub-Jovian planets ($10 \Mearth \leq m_p \leq 100 \Mearth$) with short orbital periods is an open problem. The paucity of these planets with periods $P \lesssim 3$ days is referred to as the ``Neptune desert'' \citep[e.g.,][]{Szabo+11,Beauge+13}. This deficit is especially striking considering the observationally favorable conditions for detecting planets with these characteristics \citep[Fig. 1, e.g.,][]{Mazeh+16}. However, the Transiting Exoplanet Survey Satellite (TESS; \citealt{TESS_Ricker15}) has recently unveiled a small but growing population of planets in the Neptune desert. Furthermore, a relative overdensity of sub-Jovian planets has been found outside the desert in the Neptune ``ridge'' ($3 \lesssim P \lesssim 6$ days), before giving way to the more moderately populated ``savanna'' ($P\gtrsim6$ days) \citep[e.g.,][]{CastroGonzalez+24}.

The twin actions of photoevaporation and tidal disruption during high-eccentricity migration are thought to deplete the Neptune desert \citep[e.g.,][]{Owen+18}. At close distances, strong X-ray and ultraviolet (XUV) irradiation may lead to atmospheric escape that reduces the planets' radii beneath the lower boundary of the desert \citep[e.g.,][]{Owen+13,Ionov+18,Bourrier+18,Owen+18}. At the high-mass end of the desert, planets sufficiently massive to resist significant photoevaporation may instead undergo Roche lobe overflow during high-eccentricity migration from initially "cold" orbits ($a \gtrsim 1$ au) to their present locations \citep[e.g.,][]{Wu+03,Fabrycky+07,Naoz+11}. Tidal disruption produces a period-dependent desert boundary, inward of which we expect planets to be destroyed \citep[e.g.,][]{Matsakos+16,Owen+18,CastroGonzalez+26}. Radius inflation from tidal heating during migration and orbital decay due to tidal dissipation raised in the host star may also play a role in both sculpting and populating the desert \citep[e.g.,][]{Hallatt+25,Hallatt+25b}. 

While previous work on the hot Neptune desert has primarily focused on the regime where planets are completely tidally disrupted and destroyed, \cite{Weldon+26} recently showed that the majority of giant planets ($m_p>100M_{\oplus}$) undergoing high-eccentricity migration may experience partial disruption via Roche lobe overflow at high eccentricities. Recent hydrodynamical simulations of this process \citep[][]{Fan+26} confirmed that giant planets with heavy-element cores may only lose a fraction of their gaseous envelope. Partial disruption elevates the hot Jupiter survival fraction over prior studies that assumed tidal disruption was completely destructive \citep[e.g.,][]{Naoz+12,Petrovich15b,Anderson+16}. \cite{Weldon+26} found that the enhanced survival fraction of giant planets brings theoretical hot Jupiter yields in agreement with observed occurrence rates, and angular momentum returned to the planetary orbit from debris accretion onto the host star may produce the ``three-day pileup'' in the orbital period distribution that has been observed \citep[e.g.,][]{Gaudi+05,Wright+09}. While \cite{Weldon+26} followed mass-losing planets down to the Saturn regime ($\sim0.3M_J$), with an average mass loss fraction of $\sim5-10\%$, their simulations show that it is possible that some Jovian planets could lose an even greater fraction of their gas envelopes (up to $\sim90$\%). Surviving Neptune-mass remnants ($\sim0.05M_J$) may undergo high-eccentricity migration to become hot Neptunes.

\begin{figure}
    \centering
    \includegraphics[width=\linewidth]{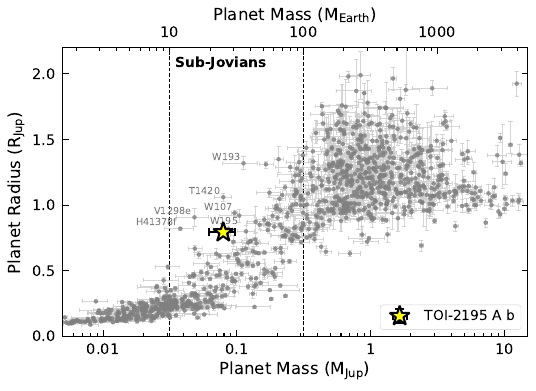}
    \includegraphics[width=\linewidth]{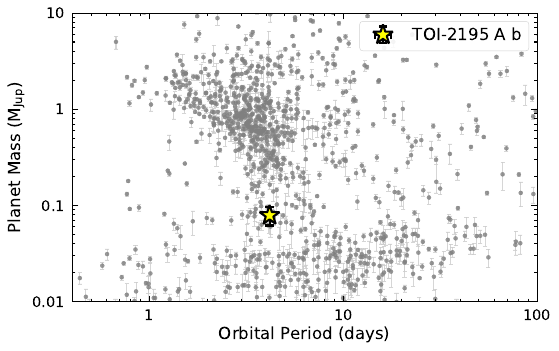}
    \caption{The newly discovered planet reported in this paper, TOI-2195\,A\,b, has a mass of 1.5\,$M_\mathrm{Nep}$, making it one of the least massive known planets with a size $> 8\Rearth$. The top panel shows the measured properties of TOI-2195\,A\,b on a mass-radius diagram, with gray points showing known planets from the NASA Exoplanet Archive \citep{PSCompPars} with radii measured to better than 20\% and RV-measured masses better than 33\%. The ``hot Neptune desert'' refers to the relative dearth of planets with masses of $0.05\,\Mjup \lesssim m_p \lesssim 0.3\,\Mjup$ on orbital periods less than three days, as seen in the mass-period diagram in the bottom panel. TOI-2195\,A\,b resides just exterior to the desert.}
    \label{fig:planet_context}
\end{figure}

Recent observational evidence has suggested a related formation pathway for at least some of the hot Neptune planets and the population of hot Jupiters. The hosts of planets in the Neptune desert and ridge show the same tendency to be metal-rich as  hot-Jupiter hosts, indicating that these populations share a common evolutionary history \citep[][]{Dong+18,Vissapragada+25}. Additionally, TESS candidate planets in the Neptune desert and hot Jupiters show an elevated stellar multiplicity rate, suggesting the potential for common dynamical processes to have aided their formation \citep[][]{EelesNolle+25}. Furthermore, the overabundance of planets in the ridge ($3 \lesssim P \lesssim 6$ days) identified by \cite{CastroGonzalez+24} resembles the three-day pileup in the hot Jupiter orbital period distribution. Additional clues come from stellar obliquities, the angles between stellar spin and planetary orbits. Many hot Jupiters exhibit significant obliquities, with some planets on misaligned and even retrograde orbits \citep[e.g.,][]{Triaud+10,Albrecht+12}. A growing number of hot sub-Jovians are also found to be on misaligned orbits \citep[e.g.,][]{Yee+25c,Bourrier+18,Espinoza-Retamal+26}. Altogether, these correlations between the properties of hot Neptunes and hot Jupiters hint at common formation channels.

Several mechanisms for inducing high-eccentricity migration may also produce the high obliquities that are observed, including planet-planet scattering \citep[e.g.,][]{Nagasawa+08,Chatterjee+08}, secular chaos in multi-planet systems \citep[e.g.,][]{Wu+11,Teyssandier+19}, or secular perturbations from a tertiary planetary or stellar companion in a hierarchical configuration \citep[e.g.,][]{Holman+97,Wu+03,Fabrycky+07,Naoz+11,Naoz+12,Petrovich15b,Petrovich+15a,Anderson+16,Weldon+24,Weldon+25}. The last of these, including octupole-level perturbations, is referred to as the Eccentric Kozai-Lidov (EKL) mechanism, a process that most readily produces the highest misalignments \citep[for a review, see][]{Naoz16}. As an alternative to the high-eccentricity migration scenario, secular resonances during protoplanetary disk dispersal may tilt the planetary orbits in systems harboring both a hot Neptune and cold Jupiter \citep[][]{Petrovich+20}. Because these scenarios require different companion configurations, observational constraints on the presence of a companion can be used to understand a planet's dynamical history.

NASA's TESS mission is discovering new sub-Jovian planets on short-period orbits, providing a growing population of such objects with which to test the theory that they are former hot Jupiters that underwent partial tidal disruption and mass loss.
TESS' all-sky survey enables the discovery of these rare planets around relatively bright stars, suitable for follow-up observations that allow us to characterize them in detail.
In this work, we report the discovery of TOI-2195~A~b, a new benchmark planet in the sub-Jovian ridge (Fig. \ref{fig:planet_context}). We confirm and characterize the planet using TESS and ground-based photometric, spectroscopic and imaging observations (\S\ref{sec:observations}--\ref{sec:system_characterization}).
TOI-2195~A~b has a remarkably low mass ($\Mp = 1.5\,\Mnep = 0.08 M_J$) despite its giant-planetary size ($\Rp = 0.8\,\Rjup$), while a Rossiter-McLaughlin measurement (\S\ref{sec:obliquity}) indicates that it likely resides on a polar orbit.
We investigate the potential formation history of this planet by modeling its inward migration via the EKL mechanism and mass loss through Roche lobe overflow (\S\ref{sec:formation}), showing that the current-day properties of TOI-2195~A~b are consistent with it forming as a Jupiter-mass object beyond the ice line and subsequently losing up to $\sim90\%$ of its mass during migration.

\section{Observations} \label{sec:observations}

In this section, we describe the TESS and ground-based follow-up observations made to confirm and characterize the transiting planet around TOI-2195~A.

\subsection{TESS Photometry} \label{ssec:tess}

TOI-2195 (TIC 24695044; $V = 11.4$~mag) was observed by NASA's Transiting Exoplanet Survey Satellite (TESS; \citealt{TESS_Ricker15}) in 11 sectors beginning with TESS' first sector of observations in 2018. The TESS Science Processing Operations Center \citep[SPOC;][]{TESS_SPOC_Jenkins2016,TESS_SPOC_Caldwell2020} conducted a transit search of Sector 27 on 28 August 2020 with an adaptive, noise-compensating matched filter \citep[][]{Jenkins+02,Jenkins+10,Jenkins+20}, producing a Threshold Crossing Event (TCE) for which an initial limb-darkened transit model was fitted \citep[][]{Li+19} and a suite of diagnostic tests were conducted to help evaluate the planetary nature of the signal \citep[][]{Twicken+18}.  The TESS Science Office (TSO) reviewed the vetting information and issued a TESS Object of Interest (TOI) alert on 18 September 2020 \citep[][]{Guerrero+21}.  The transit signature passed all the diagnostic tests presented in the Data Validation reports. For Sector 27, the host star was found to be located within $0.80 \pm 2.49''$ of the source of the transit signal. The transit signature has also been identified in searches of Full Frame Image (FFI) data by the Quick Look Pipeline (QLP) at MIT \citep[][]{Huang+20,Huang+20b}. \cite{Melton+24} independently identified the candidate as a transiting planet candidate in the DIAmante TESS AutoRegressive Planet Search (DTARPS).

We downloaded the TESS light curves from the Mikulski Archive for Space Telescopes (MAST) with the \texttt{lightkurve} package \citep{Lightkurve18}.
We elected to use the light curves produced by the TESS SPOC.
In Sectors 1, 2, and 13, TOI-2195 was only observed in the FFIs at 30-min cadence; while in later sectors following its identification as a planet candidate, it was selected for short-cadence observations (Sectors 27--29, 67--69, 94, and 96).

The SPOC pipeline automatically corrects light curves for flux contamination in the photometric aperture from nearby stars in the TESS Input Catalog \citep[TIC;][]{TIC_Stassun2018,TIC_Stassun2019}.
In the case of TOI-2195, this contamination is almost entirely due to a faint, $\Delta G = 4.2$~mag, star separated by $3\farcs3$ from the primary (see \S\ref{ssec:stellar_companion} for more discussion of this companion); there are no other bright stars in or near the photometric aperture
(Fig. \ref{fig:toi2195_tpf}).
To allow for possible inaccuracies in the dilution correction performed by the SPOC pipeline, which is based on their TIC catalog magnitudes, we reversed the automatic correction using the CROWDSAP and FLFRCSAP keywords reported in the FITS files, and allowed this dilution to be corrected as part of our global system modeling (\S\ref{ssec:system_modeling}).
In our subsequent analysis, we retained only the data around the transit events (including one transit duration pre- and post-transit), which we extracted from the light curves after detrending them using a basis spline fit to the out-of-transit continuum \citep{Keplerspline_Vanderburg2014,Keplerspline_Shallue2018}.

\subsection{Ground-Based Photometry} \label{ssec:groundphot}

Following its alert as a transiting planet candidate, we began ground-based follow-up observations of TOI-2195.01. 
We observed the transit events with seeing-limited time-series photometry with the 1.0m telescopes at the South African Astronomical observatory (SAAO) and Siding Springs Observatory (SSO) nodes of the Las Cumbres Observatory Global Telescope \citep[LCOGT;][]{LCOGT_Brown2013}, as well as the 0.36m telescope at the private El Sauce Observatory.
These observations were coordinated by the TESS Follow-up Observing Program Sub-Group 1 \citep[TFOP SG1;][]{TFOP_Collins2018} and scheduled using the \texttt{TAPIR} software \citep{TAPIR_Jensen2013}.
Data reduction and aperture photometry were performed with the \texttt{AstroImageJ} package \citep{AstroImageJ_Collins17}.

We ruled out significant chromaticity in the transit depths, as the observations from SSO were made in alternating $B$ and $z^\prime$ filters.
Additionally, small aperture reductions of these observations confirmed that the transits are indeed around the brighter star, ruling out the possibility that the events seen by TESS are due to deep eclipses on the faint companion diluted by the light of the primary.
However, because the two stars are barely resolved in these observations, in our final analysis of these data we used lightcurves derived from photometric apertures that fully contain the light of both stars, and accounted for this blending by fitting for a flux dilution parameter.

\subsection{High Angular Resolution Imaging} \label{ssec:imaging}

We checked for the presence of even closer companions using speckle imaging, which allows observations at even higher angular resolution.
% (Figure \ref{fig:toi2195_imaging}).
We observed TOI-2195 using the High Resolution Camera \citep[HRCam][]{SOAR_Tokovinin2008} mounted on the Southern Astrophysical Research (SOAR) 4.1~m telescope. The observation and data reduction strategy for the SOAR HRCam observations are described in \citealt{SOAR_TESS_Ziegler2019,SOAR_TESS_Ziegler2021,SOAR_Tokovinin2018}.
We also used the Zorro speckle imager on Gemini-South \citep{Gemini_Zorro_Alopeke_Scott2021}, with data reduction performed as in \citep{Speckle_Reduction_Howell2011}.
Neither of these observations revealed the presence of any stars closer than the known $3\farcs3$ companion, down to the instrumental sensitivity limits.
The reconstructed images and detection limits are shown in Figure \ref{fig:toi2195_imaging} in Appendix \ref{app:imaging}.

\subsection{High-Resolution Spectroscopy} \label{ssec:spectroscopy}

We obtained high-resolution spectroscopic observations of TOI-2195 to characterize the orbital properties of the planet candidate and the atmospheric properties of the host star.

\subsubsection{CHIRON Spectroscopy}
We first observed TOI-2195 using the CTIO High Resolution Spectrometer (CHIRON; \citealt{CHIRON_Tokovinin2013,CHIRON_Paredes2021}) on the SMARTS–GSU 1.5-meter Telescope at Cerro Tololo Inter-American Observatory.
We acquired ten stellar spectra of TOI-2195 between 30 Aug 2021 and 19 Nov 2022 with exposure times of 1500--1800s in image slicer mode ($R\sim 80{,}000$).
Each observation was bracketed with observations of a ThAr lamp to enable wavelength calibration.

The CHIRON data were reduced using the standard pipeline described in \cite{CHIRON_Paredes2021}.
We performed a least-squares deconvolution between each observed spectra and a spectral template from the ATLAS9 spectral library \citep{ATLAS_Castelli2003} to measure radial velocities (RVs), following the procedure in \citep{CHIRON_Zhou2020}.
The reduced spectra showed no evidence for the presence of a spectroscopic binary, while the RV measurements indicated no significant RV variation within the measurement uncertainties, which ranged from 20--60~\ms.

\subsubsection{Magellan/PFS Spectroscopy}
We subsequently observed TOI-2195 with the Planet Finder Spectrograph \citep[PFS;][]{PFS_Crane2006,PFS_Crane2008,PFS_Crane2010} on the Magellan Clay 6.5m telescope at Las Campanas Observatory, to obtain RVs with much higher precision.
The PFS observations had exposure times between 480 and 600s, and were made with an iodine absorption cell in the optical path.
We also obtained a higher S/N iodine-free template spectrum for the purpose of deriving RVs.
These data were reduced and precise RVs extracted using the pipeline from \citet{PFS_Butler1996}.
We obtained a total of 34 PFS spectra between 24 Aug 2023 and 11 Nov 2025, including 22 observations as part of a spectroscopic sequence covering the planet's transit on the night of 03 Sep 2025.
We achieved typical instrumental RV precisions of 2--3~\ms, and provide all RV measurements in a machine-readable table accompanying this paper.

\section{Analysis} \label{sec:system_characterization}

\subsection{Stellar Companion} \label{ssec:stellar_companion}

% \begin{deluxetable}{lcc}
% \tablecaption{Catalog Astrometry and Photometry for TOI-2195 System \label{tab:comp_props}}
% \tablehead{
%     & \colhead{\textbf{TOI-2195\,A}} & \colhead{\textbf{TOI-2195\,B}}
% }
% \startdata
% \input{toi2195_gaia_comp_props}
% \enddata
% \tablerefs{\tablenotemark{a}Gaia DR3 \citep{GaiaEDR3_Brown2021,GaiaEDR3_Riello2021,GaiaEDR3_Lindegren2021,GaiaDR3_RVs_Katz2022};
% \tablenotemark{b}Wide stellar binary catalog from \citet{GaiaEDR3_Binaries_El-Badry2021};
% \tablenotemark{c}2MASS \citep{TMASS_Cutri2003,TMASS_Skrutskie2006};
% \tablenotemark{d}WISE \citep{WISE_Cutri2012}.
% }
% \end{deluxetable}

As noted previously, TOI-2195 is accompanied by a fainter star with a sky separation of $3\farcs3$. 
Astrometry from the third Data Release (DR3) of the \textit{Gaia} mission \citep{GaiaEDR3_Brown2021,GaiaDR3_Vallenari2022} indicates that the two stars have similar parallaxes, proper motions, and radial velocities, suggesting that they are likely a bound pair with a projected separation of 580~AU.
\citet{GaiaEDR3_Binaries_El-Badry2021} also identified the two stars as a bound binary system based on their parallax and proper motions (the RV of the secondary was not available in \textit{Gaia} eDR3 which was used to construct their catalog), with a low chance alignment probability $\mathcal{R} = 3\times10^{-6}$.
Given that the two stars are likely part of the same stellar system, we subsequently refer to the brighter, planet-hosting star as TOI-2195\,A and the secondary as TOI-2195\,B.

\subsection{Spectroscopic Stellar Characterization} \label{ssec:spec_char}

We measured stellar atmospheric parameters for TOI-2195\,A by applying the \texttt{SpecMatch-Empirical} code \citep{SpecMatchEmp_Yee2017} to the iodine-free PFS template spectrum.
In brief, \texttt{SpecMatch-Empirical} compares the target spectrum to a library of empirical high-resolution spectra of stars with well-determined fundamental properties.
The final derived parameters are obtained from a linear combination of the stellar properties of the five stars whose spectra best match the target.
This analysis found that TOI-2195~A is an early K-type star ($\Teff = 5160 \pm 110\,K$, $\Rstar = 0.90\pm0.09\,\Rsun$) with an enhanced metallicity of $\feh = +0.26 \pm 0.08$~dex, consistent with the finding of \cite{Vissapragada+25} that hot Neptunes, just like hot Jupiters, are preferentially found around metal-rich stars.

\subsection{Global System Modeling} \label{ssec:system_modeling}

\begin{figure*}
    \centering
    \includegraphics[width=0.8\linewidth]{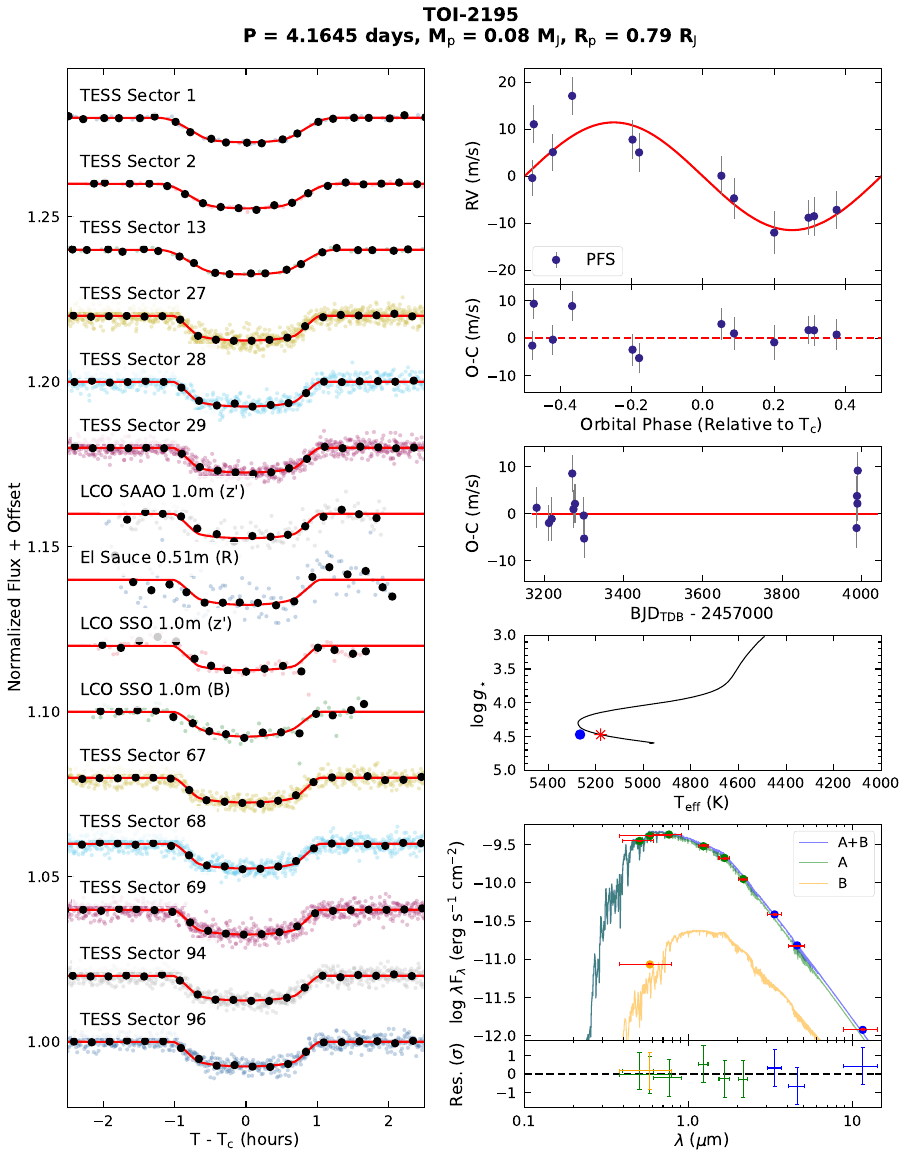}
    \caption{Left: Phase-folded transit photometry from TESS and ground-based facilities. Black circles show the data binned to a 15-minute cadence. The best-fit transit model is shown as the red line. Right, from top to bottom: (a) Phase-folded PFS RVs, excluding those taken as part of the RM sequence. The best-fit Keplerian RV model is plotted in red. (b) Residual RV time series after subtracting the best-fit model for TOI-2195\,A\,b. (c) MIST stellar evolution track (black line), with the point corresponding to the best-fit stellar age is marked with the red asterisk. The blue circle shows the best-fit stellar \Teff and \logg; the discrepancies between the two are within the fitted uncertainties. (d) Observed catalog broadband photometry for the TOI-2195\,AB system (red; horizontal error bars correspond to the width of the photometric bandpass). The circles show the model flux from the MIST bolometric correction grid given the best-fit stellar properties. We also plot atmospheric models from \citet{Kurucz1993} for illustrative purposes, but the fit is performed directly to the MIST grid. The data used to create this figure are available in the online article.}
    \label{fig:global_analysis}
\end{figure*}

We modeled the physical properties of the transiting planet, its stellar host TOI-2195\,A, and the secondary star in the system with the \Exofast code \citep{ExoFAST_Eastman2013,ExoFASTv2_Eastman19}, following the general procedures outlined in \citet{Yee+25}.
The two stars are constrained by fitting observed broadband fluxes to those derived from MIST evolutionary models \citep{MIST0_Dotter2016,MISTI_Choi2016}.
We used catalog $G$, $G_\mathrm{BP}$, and $G_\mathrm{RP}$ photometry from \textit{Gaia} DR3 \citep{GaiaEDR3_Riello2021},\footnote{We only used the $G$-band photometry for the secondary star, as we found its $G_\mathrm{BP}$, and $G_\mathrm{RP}$ magnitudes to be potentially unreliable.} JHK photometry of the primary from 2MASS \citep{TMASS_Cutri2003,TMASS_Skrutskie2006}, and W1--3 photometry from WISE \citep{WISE_Cutri2012}, where we assumed that the WISE photometry is a blend of both stars.
We placed Gaussian priors on the distances of both stars based on their \textit{Gaia} parallax measurements, as well as on the current metallicity of TOI-2195\,A from the spectroscopic measurement of \feh described in \S\ref{ssec:spec_char}.
We required the two stars to be coeval and have the same age, initial \feh, and extinction, where we placed an upper limit on their common $V$-band extinction from the extinction maps of \citet{Schlafly2011}.

To constrain the planet, we fitted the TESS and ground-based time-series photometry.
To account for dilution in the apparent transit depth arising from the light of TOI-2195\,B, we used \Exofast to compute the expected flux ratios of the two stars in each photometric band and used these to perform a self-consistent correction of each light curve.
We also fitted the PFS RVs, although we excluded the CHIRON RVs given their low precision compared with the final RV semi-amplitude.
A preliminary fit revealed no statistically significant long-term RV trend in either the PFS or CHIRON RV measurements, so we did not include any linear trends in the final model.

We sampled the posterior probability distributions and derived uncertainties with the Differential Evolution-
Monte Carlo Markov Chain (DE-MCMC) code in \Exofast.
We used a convergence criterion of $>1{,}000$ independent draws and a Gelman-Rubin statistic $\mathrm{GR} < 1.01$ for the MCMC exploration.
Table \ref{tab:exofast_results} presents the best-fit results and 68\% confidence intervals for the stellar and planetary parameters.

With a planetary mass of $m_p = 0.079 \pm 0.017 \,\Mjup = 25.1 \pm 5.4\,\Mearth$ but a radius of $\Rp = 0.79\pm0.02\,\Rjup = 8.9\pm0.2\,\Rearth$, TOI-2195\,A\,b stands out as one of the least massive planets with a giant planetary size (Fig \ref{fig:planet_context}).
This places TOI-2195\,A\,b amongst the ``super-puff'' Neptunes or ``popcorn planets'', which \citet{Yee2026} defined as objects with masses of $0.05\,\Mjup<m_p<0.2\,\Mjup$ and low bulk densities $\rhop < 0.3\,\gcc$.
This category includes other well-studied Neptunes like WASP-107\,b, TOI-1420\,b, and TOI-1173\,A\,b, the last of which has a mass and radius almost identical to TOI-2195\,A\,b ($\Rp = 0.78\pm0.02\,\Rjup$, $m_p = 0.089\pm0.03\,\Mjup$).
The inflated radii of these planets pose a challenge to standard planet formation theories, as structure models in hydrostatic equilibrium require these planets to have core masses $\lesssim 5\,\Mearth$, thought to be too low to initiate runaway gas accretion \citep[e.g.,][]{Piaulet2021,Marston2026}.
An alternative explanation is that these planets may be inflated by mechanisms like tidal heating \citep{Millholland+20} or Ohmic dissipation \citep{Batygin2025}, or that dusty outflows inflate the observed transit radius \citep[][]{Wang+19}. In the case of TOI-2195\,A\,b, we show in \S\ref{sec:neptunemigration} that it may still be contracting following significant radius inflation during a high-eccentricity phase of its migration.

\begin{deluxetable*}{lccc} \label{tab:exofast_results}
\tablecaption{Median values and 68\% Confidence Intervals for \Exofast fit of TOI-2195}
\tablehead{\colhead{Parameter} & \colhead{Description} & \multicolumn{2}{c}{Value}}
\startdata
\\[-\normalbaselineskip]\multicolumn{2}{l}{Stellar Parameters:} & TOI-2195\,A & TOI-2195\,B \\
~~~~$M_\star$ ($M_\odot$) & Stellar mass & $0.905^{+0.036}_{-0.028}$ & $0.478^{+0.031}_{-0.032}$ \\
~~~~$R_\star$ ($R_\odot$) & Stellar radius & $0.916^{+0.027}_{-0.024}$ & $0.457^{+0.025}_{-0.024}$ \\
~~~~$\log{g_\star}$ (cgs) & Stellar surface gravity & $4.471^{+0.026}_{-0.027}$ & $4.798^{+0.033}_{-0.032}$ \\
~~~~$\rho_\star$ (g cm$^{-3}$) & Stellar density & $1.66 \pm 0.14$ & $7.07^{+0.93}_{-0.79}$ \\
~~~~$L_\star$ ($L_\odot$) & Stellar luminosity & $0.582 \pm 0.018$ & $0.0299 \pm 0.0018$ \\
~~~~$T_\mathrm{eff}$ (K) & Stellar effective temperature & $5266^{+72}_{-76}$ & $3552^{+65}_{-64}$ \\
~~~~$[\mathrm{Fe/H}]$ (dex) & Metallicity & $0.309^{+0.075}_{-0.078}$ & $0.360^{+0.089}_{-0.094}$ \\
~~~~$[\mathrm{Fe/H}]_0$ (dex)\tablenotemark{a} & Initial metallicity & $0.312^{+0.071}_{-0.073}$ & $0.312^{+0.071}_{-0.073}$ \\
~~~~Age (Gyr)\tablenotemark{a} & Stellar age & $9.8^{+2.5}_{-3.4}$ & $9.8^{+2.5}_{-3.3}$ \\
~~~~EEP & Equal evolutionary phase & $379^{+16}_{-28}$ & $312^{+7}_{-12}$ \\
~~~~$A_V$ (mag)\tablenotemark{a} & Visual extinction & $0.052^{+0.031}_{-0.033}$ & $0.052^{+0.031}_{-0.033}$ \\
~~~~d (pc)\tablenotemark{a} & Distance & $174.67 \pm 0.30$ & $174.67 \pm 0.30$ \\
\\[-\normalbaselineskip]\multicolumn{2}{l}{Planet Parameters:} & TOI-2195\,A\,b \\
~~~~$P$ (days) & Period & $4.16451307 \pm 0.00000054$ \\
~~~~$T_c$ (BJD$_\mathrm{TDB}$) & Time of conjunction & $2459580.84798 \pm 0.00012$ \\
~~~~$R_p$ ($R_\mathrm{J}$) & Planet radius & $0.793^{+0.024}_{-0.022}$ \\
~~~~$m_p$ ($M_\mathrm{J}$) & Planet mass & $0.079 \pm 0.017$ \\
~~~~$\left(R_p / R_\star\right)^2$ & Planet-star area ratio & $0.007898^{+0.000094}_{-0.000098}$ \\
~~~~$K$ (m/s) & RV semi-amplitude & $10.6^{+2.2}_{-2.3}$ \\
~~~~$a$ (AU) & Semimajor axis & $0.04901^{+0.00065}_{-0.00052}$ \\
~~~~$a/R_\star$ & Planet-star separation & $11.51^{+0.31}_{-0.34}$ \\
~~~~$i$ (deg) & Inclination & $86.07^{+0.20}_{-0.26}$ \\
~~~~$b \equiv a\cos{i}/R_\star$ & Transit impact parameter & $0.770^{+0.010}_{-0.012}$ \\
~~~~$e$ & Eccentricity & $0.060^{+0.065}_{-0.042}$ \\
~~~~$\omega$ (deg) & Argument of periastron & $152^{+42}_{-83}$ \\
~~~~$\sqrt{e}\cos{\omega}$ & Eccentricity vector & $-0.14^{+0.22}_{-0.18}$ \\
~~~~$\sqrt{e}\sin{\omega}$ & Eccentricity vector & $0.08^{+0.12}_{-0.13}$ \\
~~~~$\rho_p$ (g cm$^{-3}$) & Planet density & $0.196^{+0.046}_{-0.045}$ \\
~~~~$\log{g_p}$ (cgs) & Planet surface gravity & $2.49^{+0.09}_{-0.11}$ \\
~~~~$T_\mathrm{eq}$ (K) & Planet equilibrium temperature & $1098 \pm 10$ \\
~~~~$\langle F \rangle$ (10$^9$ erg s$^{-1}$ cm$^{-2}$) & Incident flux & $0.327 \pm 0.012$ \\
~~~~$T_{14}$ (days) & Transit duration & $0.08698^{+0.00067}_{-0.00068}$ \\
~~~~$\tau$ (days) & Ingress/egress duration & $0.01598^{+0.00076}_{-0.00078}$%
\enddata

\tablenotetext{a}{These parameters are linked for both stellar components.}
\end{deluxetable*}

\section{Stellar Obliquity} \label{sec:obliquity}

We measure the projected stellar obliquity $\lambda$ using the PFS RV time series from the night of UT 03 Sep 2025 with the \texttt{rmfit} code \citep[][]{Stefansson+22}. This code models the Rossiter-McLaughlin (RM) anomaly using the formalism of \cite{Hirano+10,Hirano+11}, using the standard parametrization of the RM parameters in terms of $\lambda$ and $v\sin i$. 

We place Gaussian priors on the planet's orbital parameters based on the output of the \Exofast analysis. For $\vsini$, we use an upper limit of $3$ km s$^{-1}$, which was determined by measuring the spectral line widths in the PFS template specrum with the \texttt{SpecMatch-Synthetic} code \citep{SpecMatchSynth_Petigura2015}. We use the exoCTK web tool \citep[][]{Bourque+21} to compute limb-darkening coefficients $u_1$ and $u_2$ and centered a Gaussian prior with width 0.1 on the calculated values. We place a Gaussian prior on $v_{\beta}$ of 3.2 km s$^{-1}$ with dispersion 1 km s$^{-1}$, based on the instrumental line profile and macroturbulence as computed using the relation of \cite{Valenti+05}. We use uninformative priors for the remaining parameters. All priors are reported in Table \ref{tab:RM_priors}.

\begin{figure*}
    \centering
    \includegraphics[width=0.9\linewidth]{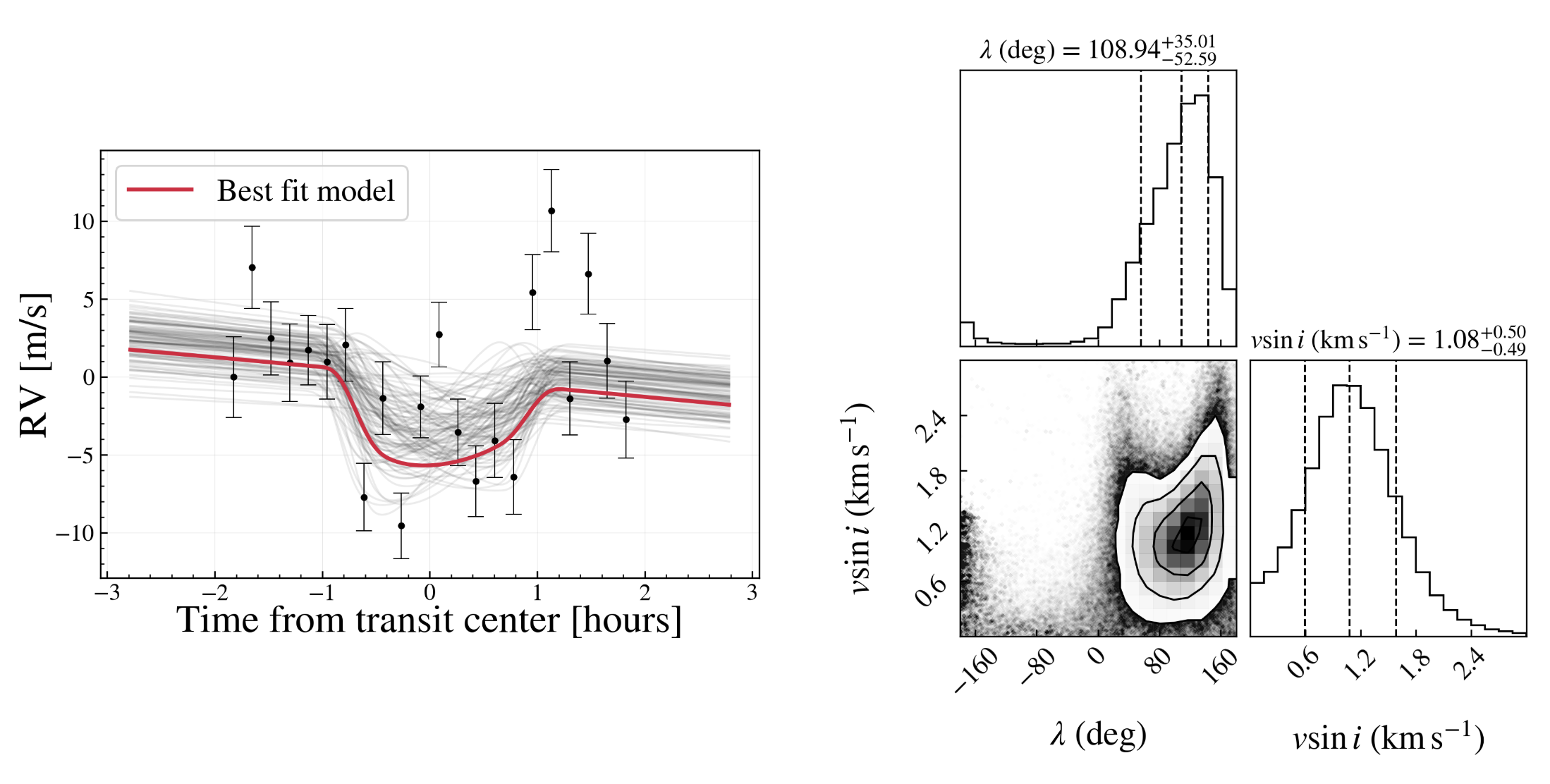}
    \caption{Left: PFS RV measurements for TOI-2195 A taken on the night of UT 03 Sep 2025. The data are shown as black points, where error bars have been inflated by the best-fit RV jitter. The red line shows the best fit RM model, and gray lines show 100 realizations of the posterior parameters. Right: Corner plot showing the posterior distributions of $\lambda$ and $v \sin i$.}
    \label{fig:rm_plot}
\end{figure*}

We use the differential evolution optimization code \citep[][]{Storn+97} implemented in \texttt{PyDE16}\footnote{https://github.com/hpparvi/PyDE} to find
the maximum-likelihood RM model. Uncertainties on model parameters are obtained by sampling the posterior probability distributions using the MCMC sampler \texttt{emcee} \citep[][]{Emcee_Foreman-Mackey13}. The best-fits for each parameter and uncertainties are reported in Table \ref{tab:RM_priors}. The best fit to the data and 100 model realization drawn from the posterior are shown in the left panel of Figure \ref{fig:rm_plot}. We obtain an obliquity measurement of $\lambda = {109^{+35}_{-53}} ^{\circ}$, indicating the planet is likely on a highly misaligned, near-polar orbit about its host star. $\lambda\leq0^{\circ}$ is excluded with 95.6\% confidence.

Due to the variable conditions on the night of the RM observations, the RV timeseries exhibits moderate jitter, leading to a tail to negative $\lambda$ that dominates the lower error bar (see right panel of Figure \ref{fig:rm_plot}). This tail is correlated with low $v\sin i$ (i.e. a scenario where the RM effect is not detected and $\lambda$ is unconstrained). We evaluate the significance of the RM signal using a bootstrap resampling test comparing the improvement of the RM model over a straight-line model. Under the null hypothesis of a linear trend plus noise, only 0.41\% (0.43\%) of trials produce an improvement in the root mean square residuals ($\Delta \chi^2$) at least as large as observed, corresponding to $\sim2.6\sigma$ significance. 
We also checked for possible transit timing variations (TTVs) by fitting the TESS data allowing for an independent transit time for each observed transit. We did not detect any statistically significant departures from a linear ephemeris, with a typical precision on the transit timing of $\pm 1$~minute per epoch, so it is unlikely that our spectroscopic timeseries missed the transit event of the planet.
Based on these considerations, we conclude that we likely detected the Rossiter-McLaughlin effect for TOI-2195\,A\,b, which indicates that the planet likely resides on a polar orbit.

\begin{deluxetable*}{lccc} \label{tab:RM_priors}
\tablecaption{Priors and Fit Results for RM analysis}
\tablehead{\colhead{Parameter} & \colhead{Prior} & \colhead{Posterior}}
\startdata
$T_c$ (BJD$_{\rm TDB}$) & $\mathcal{N}$(2459580.84798,0.00012) & $2459580.84798\pm 0.00012$\\
$P$ (days) & $\mathcal{N}$(4.16451307,0.00000054)& $4.16451308^{+0.0000006}_{-0.0000005}$\\
$\lambda$ (deg) & $\mathcal{U}$(-180.0,180.0)& 108.94$^{+35.01}_{-52.59}$\\
$v \sin i_*$ (km/s) & $\mathcal{U}$(0,3) & 1.08$^{+0.50}_{-0.49}$\\
$i_*$ (deg) & $\mathcal{N}$(86.07,0.26)&$86.06\pm0.24$\\
$R_p/R_*$ & $\mathcal{N}$(0.08887,0.00055)&$0.08886\pm0.00059$\\
$a/R_*$ & $\mathcal{N}$(11.51,0.34)&$11.52\pm0.32$\\
$u_1$ & $\mathcal{N}$(0.66,0.10)&$0.659\pm0.099$\\
$u_2$ & $\mathcal{N}$(0.10,0.10)&0.098$^{+0.100}_{-0.099}$\\
$e$ & $\mathcal{N}$(0.06,0.060)&$0.060^{+0.059}_{-0.058}$\\
$\omega$ & $\mathcal{U}$(0,360)&190.0$^{+130.0}_{-150.0}$\\
$v_{\beta}$ (km/s) & $\mathcal{N}$(3.2,1.0)&3.22$^{+0.98}_{-0.97}$\\
$K$ (m/s) & $\mathcal{N}$(10.6,2.3)&$10.3^{+2.3}_{-2.2}$\\
$\gamma$ (km/s) & $\mathcal{N}$(-0.9,1.6)&0.5$^{+1.1}_{-1.2}$\\
$\sigma_{RV}$ (m/s) & $\mathcal{U}$(0,10) &4.0$^{+1.0}_{-0.85}$
\enddata
\end{deluxetable*}

\section{Formation History} \label{sec:formation}

The presence of a wide stellar binary companion in the TOI-2195 system and the potentially misaligned orbit of the planet suggest that EKL-driven high-eccentricity migration may have played a role in its formation. The low planetary density also suggests that an active dynamical history contributed to radius inflation. In this Section, we investigate the formation of the planet from high-eccentricity migration considering two progenitor scenarios. First, we explore the formation of the planet as an original Neptune. Then, we consider the possibility that the planet started as a higher mass Jovian that underwent mass loss during the migration process.

\subsection{Migration as an original Neptune}

\label{sec:neptunemigration}

\begin{figure}
    \centering
    \includegraphics[width=\linewidth]{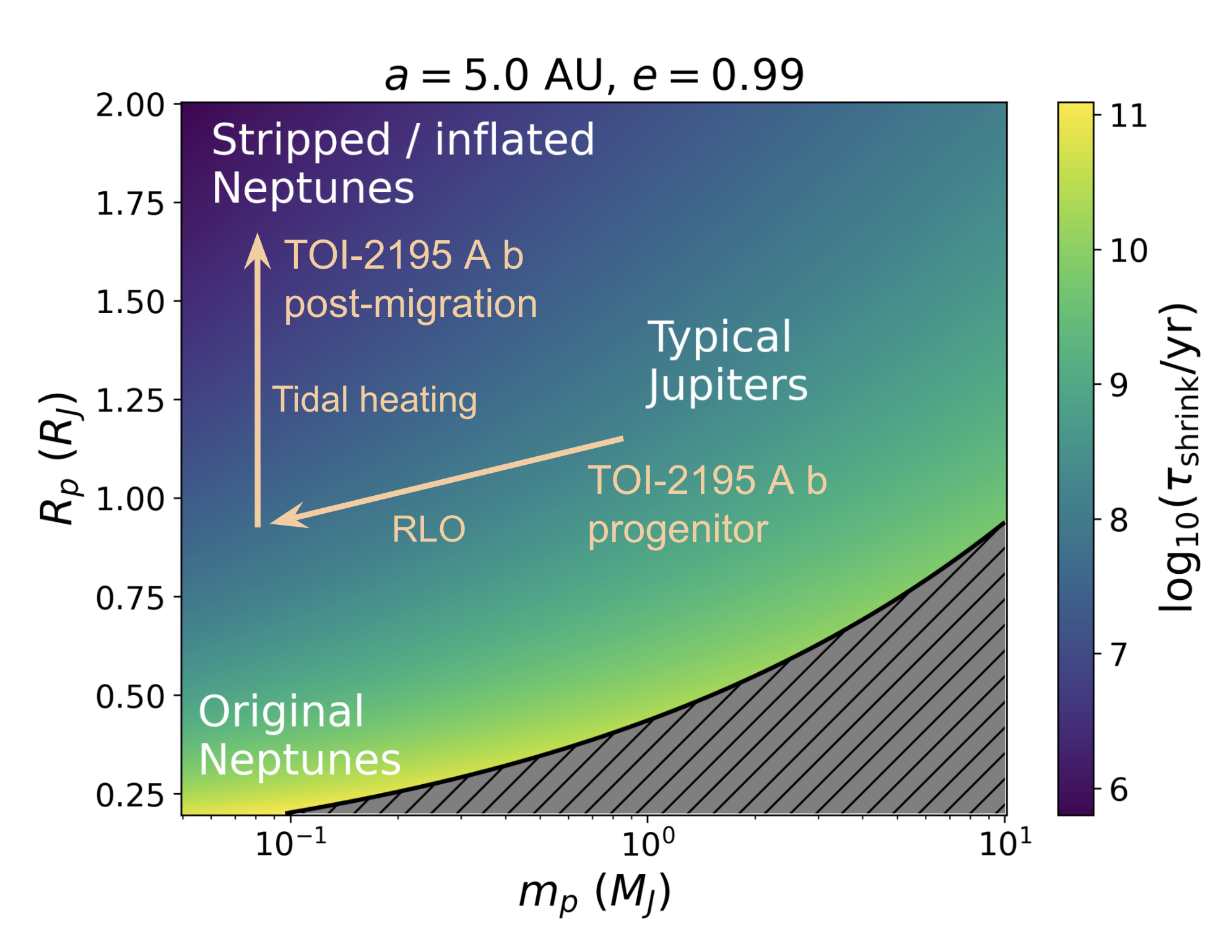}
    \caption{Timescale (color code) for tides to shrink a planetary orbit (see Eq. {\ref{eq:shrinkingtime}}), as a function of varied planetary radius and mass. We consider a characteristic distant, eccentric orbit ($a=5 $ au, $e=0.99$) for planets undergoing high-eccentricity migration and set $k_{2,p}=0.5$ and $Q_p=10^5$. We overplot a potential formation history for TOI-2195 A b. The gray shaded region corresponds to densities $>15$ g cm$^{-3}$, beyond which we generally do not expect to find planets \citep[e.g.,][]{Guenther+13}.}
    \label{fig:shrinkingtime}
\end{figure}

At first, it may seem unlikely to form a hot Neptune from conventional high-eccentricity tidal migration if the planet initially started with a Neptune-like mass and radius, as the timescale for tidal migration can be longer than a Hubble time. Indeed, this issue has been noted before \citep[e.g.,][]{Giacalone+17,Espinoza+26}. In Figure \ref{fig:shrinkingtime}, we calculate the timescale for tides to shrink the orbit \citep{Hut81}
\begin{equation}
    t_{\rm shrink} = \frac{4}{21} \frac{Q_p}{k_{2,p}n_p} \frac{m_p}{M_*} \left(\frac{a_p}{R_p}\right)^5 \frac{(1-e^2)^{15/2}}{e^2f(e)} \, ,
\end{equation}
where $Q_p$ is the tidal quality factor of the planet, $k_{2,p}$ is the planetary Love number, $n_p$ is the mean motion, and
\begin{equation}
    f(e) = \frac{
1 + \frac{45}{14}e^2 + 8e^4 + \frac{685}{224}e^6 + \frac{255}{448}e^8 + \frac{25}{1792}e^{10}
}{
1 + 3e^2 + \frac{3}{8}e^4
} \ .
\label{eq:shrinkingtime}
\end{equation}
We consider characteristic values for a distant ($a_p=5$ au), highly eccentric ($e_p=0.99$) planet undergoing high-eccentricity migration with $k_{2,p}=0.5$ and $Q_p=10^5$. Here, we assume $Q_p$ is the same for Jupiters and Neptunes. This assumption is motivated by studies of the Solar System \citep[e.g.,][]{Goldreich+66,Wang+25} and extrasolar planets \citep[e.g.,][]{Essick+16,Louden+24}, which tend to find values of $Q_p\sim10^5$. Given our limited understanding of tidal dissipation and varying estimates in the literature, it is possible that $Q_p$ differs across the planetary mass spectrum, widening the parameter space for tidal migration. Additionally, the shrinking timescale depends sensitively on the choice of planetary orbit, which varies in time and may in some cases reach a sufficiently close periapse to enable tidal migration for a subset of the population. However, increasing the eccentricity also increases the possibility for tidal disruption, and the window to migrate in time without being disrupted is narrow \cite[][]{CastroGonzalez+26}. Here, we choose characteristic values to heuristically illustrate the problem of long migration for most hot Neptunes \citep[e.g.,][]{Espinoza+26}, and we will explore the population in more detail below.

Examining Figure \ref{fig:shrinkingtime}, we see a range of shrinking timescales, as $t_{\rm shrink} \propto m_p/R_p^5$. Typical Jupiter-like planets undergo tidal migration to form hot Jupiters on timescales of $\sim 10^8-10^9$ yr. Planets with initially Neptune-like radii and masses (lower left corner) undergo tidal migration on timescales $>10^{10}$ yr, indicating that initially cold Neptunes are less likely to migrate on observable timescales. Planets may instead begin as giant planets that overfill their Roche lobes at high eccentricities \citep[][]{Weldon+26}. Neptune-mass objects with Jupiter-like radii arising from tidal mass loss undergo very rapid migration on timescales $\lesssim 10^6-10^7$ yr that is enhanced by radius inflation due to tidal heating during the migration. We overplot in Figure \ref{fig:shrinkingtime} a potential formation history for TOI-2195 A b, starting as a Jupiter-like object and ultimately ending up as an inflated Neptune. Both the increased radius and the reduced mass allow tides to act efficiently form a hot planet. This effect motivates us to consider migration as a stripped giant in Section \ref{sec:jupitermigration}.

To assess the likelihood of forming a TOI-2195-like system in more detail starting as an original Neptune, we perform numerical simulations for a population of 400 systems consisting of a host star, initially cold Neptune, and a binary stellar companion. We numerically evolve the systems by solving the hierarchical three-body equations of motion, including the effects of general relativity, tidal friction in both the planet and star, and stellar evolution (see Appendix \ref{app:numericalmethods} for detailed methods). In Section \ref{sec:jupitermigration}, we will include the effects of planetary structure, including tidal mass loss and tidal heating, to investigate formation as an original Jovian. We do not yet include these effects, as we are interested in exploring whether the migration of fixed-mass Neptunes is possible. We run each simulation for 5 Gyr, and stop the evolution if a planet crosses $\eta r_t$, where
\begin{equation}
    r_t = R_p \left(\frac{M_*+m_p}{m_p}\right)^{1/3} \, ,
\end{equation}
is the Roche limit. We set $\eta=2$, which corresponds roughly to the boundary where hydrodynamical simulations show planets undergo significant disruption \citep[e.g.,][]{Faber+05,Guillochon+11,Fan+26}.

For the initial conditions, observationally constrained parameters are fixed using approximate values: the companion semi-major axis $a_c = 600$ au\footnote{We have chosen to fix $a_c$ as the approximate projected separation of the stellar companion. If the binary orbit is eccentric, it is likely that the true semimajor axis is smaller given the higher likelihood of observing the star near apoapse. This choice is conservative, as a smaller semimajor axis would increase the strength of EKL perturbations. This would widen the parameter space for EKL to act and lead to a higher production rate of hot Neptune formation at the population level.}, the host star mass $M_* = 0.9 M_{\odot}$, and companion mass $m_c = 0.5 M_{\odot}$, while other parameters are drawn using a Monte Carlo approach. The planet is initialized uniformly between $a_p = 5-10$ au, planetary eccentricity $e_p$ drawn from a Rayleigh distribution with mean 0.1, companion eccentricity $e_c$ drawn uniformly from 0-1, the mutual inclination $i$ between the inner and outer orbits sampled isotropically, the arguments of periapse $g_1$ ($g_2$) of the inner (outer) orbit sampled uniformly from $0-360^{\circ}$, and stellar obliquity $\psi$ with a Rayleigh distribution with a mean of $10^{\circ}$. The small amounts of initial eccentricity and obliquity are consistent with gentle early interactions between the planet and any other planets, the protoplanetary disk, and stellar companion \citep[e.g.,][]{Beauge+12,Ragusa+18,Li+23,Lu+25,Weldon+25}. We fix the mass of the planet $m_p$ to $0.08M_J$ and consider planetary radii $R_p$ drawn uniformly from $0.25-0.5R_J$. 

In our simulations, $25.8\%$ of systems are tidally disrupted and $5.8\%$ of systems migrate and survive as a hot Neptunes. We examine the final semimajor axes of the migrated planets and find that none of the systems are compatible with the 0.049 au semimajor axis of TOI-2195 A b. The planets in the simulations have semimajor axes ranging from 0.019 to 0.041 au, with a median of 0.025 au, indicating that the production of TOI-2195 A b from this process is unlikely. The distribution of simulated planetary periods falls short of the broader population of Neptune ridge planets with orbital periods $P\sim 3-6$ days \citep[][]{CastroGonzalez+24}.

We can also demonstrate this discrepancy from an analytical understanding of Roche lobe overflow during migration. Migrating planets that reach high eccentricities follow tracks of nearly constant angular momentum to reach final semimajor axes \citep[e.g.,][]{Faber+05}
\begin{equation}
    a_f = a(1-e^2) \sim 2a(1-e) = 2q \, ,
\end{equation}
where $q$ is the periastron. Therefore, $a_f \sim 2 \eta r_t$ for the small fraction of planets that avoid disruption but migrate in 5 Gyr. For an original Neptune ($m_p = 0.08 R_J$, $R_p = 0.4R_J$) with $\eta=2$ fixed, this gives $a_f \sim 0.02$ au. This distance falls significantly short of TOI-2195 A b, which lies outward in the Neptune ridge at $a_p \sim 0.05$ au. If we consider instead that the planet lost mass as an original Jovian (reaching $m_p = 0.08M_J$, but maintaining $R_p \sim 1.0R_J$), then the threshold for the final location moves outward to $a_f \sim 0.04$ au. Additional structural and dynamical considerations may push this location further outward. This motivates us to consider the full structural evolution of original giant planets that undergo mass loss to become hot Neptunes.

\begin{figure*}
    \centering
    \includegraphics[width=\linewidth]{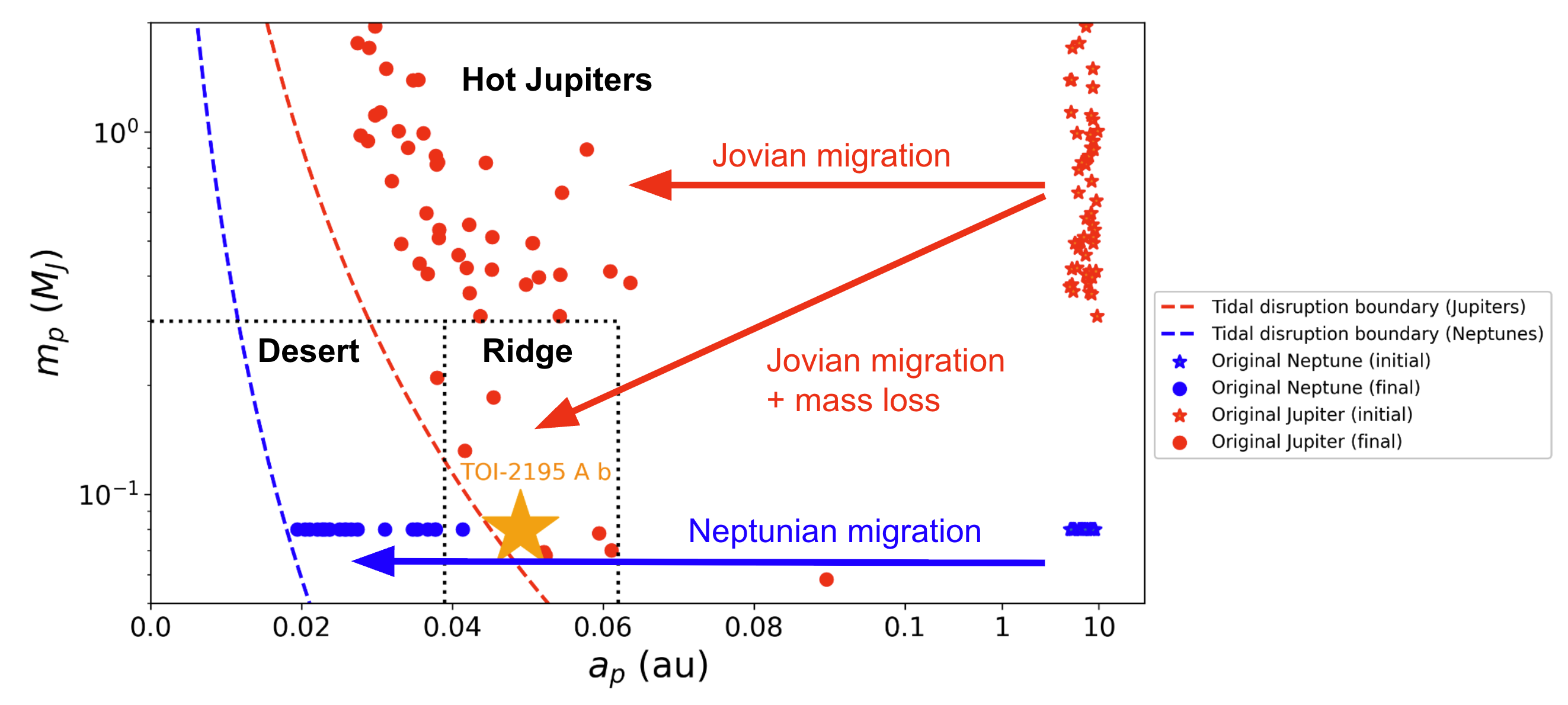}
    \caption{Initial and final masses and semi-major axes of simulated systems. Red stars correspond to the initial conditions for original Jupiters, and red dots correspond to the final conditions for original Jupiters. Blue stars correspond to the initial conditions for original Neptunes, and blue dots correspond to the final conditions for original Neptunes. TOI-2195 A b is marked with an orange star. The analytic tidal disruption boundaries are shown for Neptunes (blue) and Jupiters (red) with dashed lines. Red arrows schematically show migration pathways for originally Jovian planets (with little mass loss and significant mass loss). The blue arrow schematically shows the migration pathway for original Neptunes. Approximate locations of the Neptune desert and ridge are labeled and outlined with dashed black lines.}
    \label{fig:migrations}
\end{figure*}

\subsection{Migration as a stripped Jupiter}

\label{sec:jupitermigration}

Beginning as a Jovian planet may widen the parameter space for migration and allow planets to migrate to larger periods in the ridge. To investigate this formation, we perform a population synthesis study with 500 systems of initial Jovians. We numerically evolve the systems by solving the hierarchical three-body system equations of motion, including the effects of general relativity, tidal dissipation in both the planet and star, planetary cooling, tidal heating of the planet, planetary mass loss, and stellar evolution (see Appendix \ref{app:numericalmethods} for detailed methods). Initial conditions are drawn using the same distributions as described in Section \ref{sec:jupitermigration}, except we now draw the planetary mass $m_p$ between $0.3-2M_J$ from a power-law distribution with $dn/dm_p \propto m_p^{-1}$ \citep[e.g.,][]{Marcy+2000}. Therefore, we consider original Saturns and Jupiters that are more massive than the present-day mass of TOI-2195 A b, and the mass and radius evolution are set by the planetary structure evolution. This work is intended to be a proof-of-concept study, and a more complete parameter space search will be explored in Weldon et al. in prep.

In Figure \ref{fig:migrations}, we show the simulation outcomes for both the systems that start as cold Jupiters and the systems that started as cold Neptunes in Section \ref{sec:neptunemigration}. TOI-2195 A b is marked with an orange star. We see that the final locations of original Neptunes (blue dots) fall short of the Neptune ridge, where TOI-2195 A b is located. Systems in the ridge generally require migration timescales longer than 5 Gyr. However, the final locations of original Jupiters (red dots) that have lost significant mass lie within the Neptune ridge. We plot the tidal disruption boundary for initial Jupiters as a dashed red line (setting $R_p =1R_J$), and the tidal disruption boundary for initial Neptunes as a dashed blue line (setting $R_p=0.4M_J$), as discussed above in Section \ref{sec:neptunemigration}. We see that these thresholds bound the simulated populations, and that TOI-2195 A b is consistent with originally Jovian planets that lose mass while maintaining a Jupiter-like radius.

\begin{figure*}
    \centering
    \includegraphics[width=\linewidth]{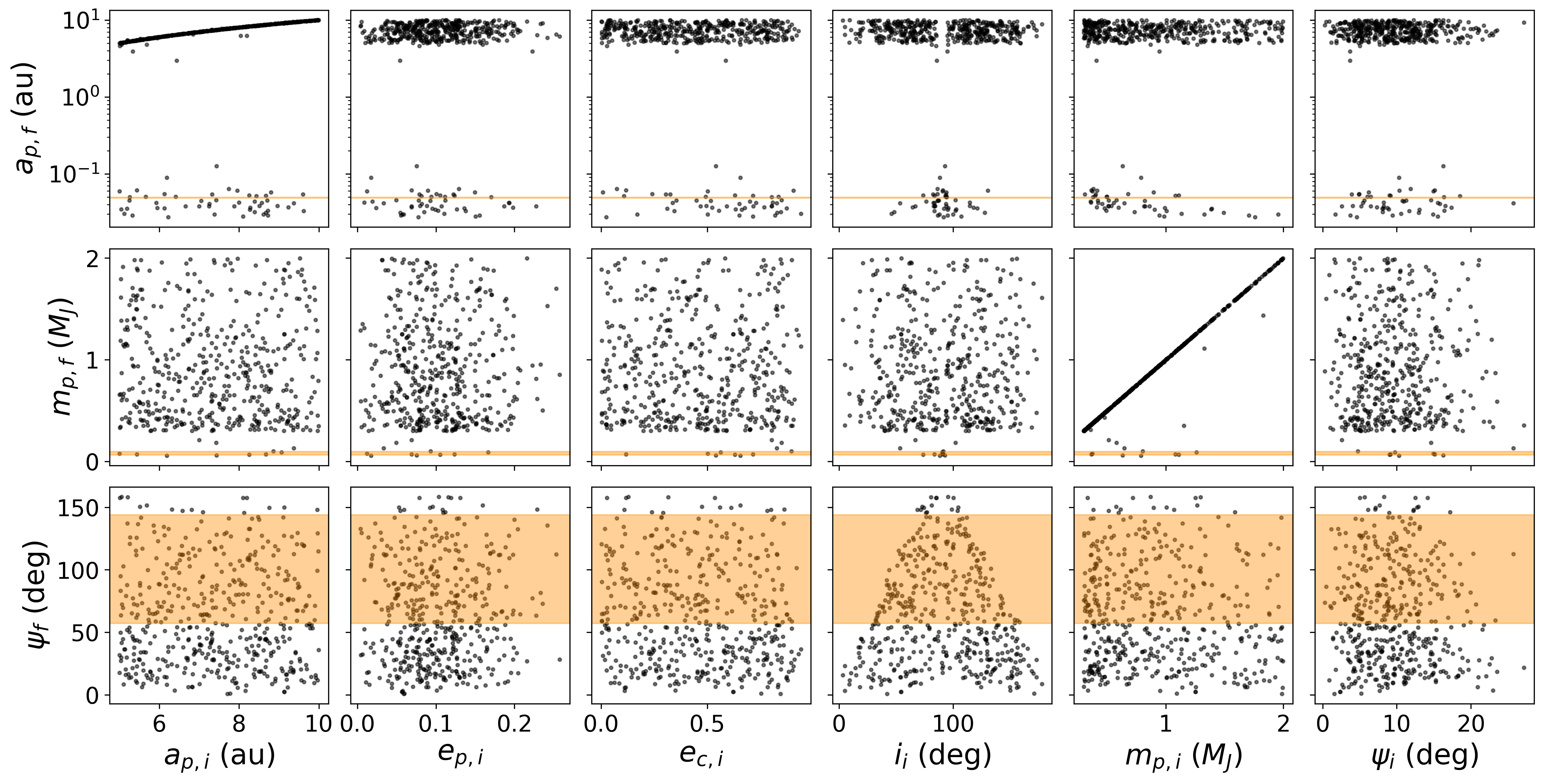}
    \caption{Final $a_p$, $m_p$, and $\psi$ of simulated systems (black dots) as a function of initial system conditions. $a_c = 600$ au, $M_* = 0.9 M_{\odot}$, and $m_3 = 0.5 M_{\odot}$ are fixed, and the other initial conditions are drawn using a Monte Carlo approach. Shaded orange regions show the observational constraints for TOI-2195 b (the sky projection $\lambda$ is shaded for the obliquity).}
    \label{fig:toi2195pop}
\end{figure*}

In Figure \ref{fig:toi2195pop}, we show the simulation outcomes (black dots) as a function of initial conditions. The shaded orange regions show the measured parameters for TOI-2195 A b. As depicted in the Figure, it is straightforward to form hot planets ($a_p<0.1$ au), stripped sub-Jovians ($m_p<0.3 M_J$), and planets on nearly polar orbits ($\psi \sim 90^{\circ}$). 9.4\% of systems survive as a hot planet (Jovian or sub-Jovian), 1.8\% of systems as a hot Neptune ($m_p < 0.3M_J$), and 1.0\% as a hot Neptune on a highly misaligned orbit ($\psi>70^{\circ}$).\footnote{We emphasize that the initial conditions are set with the parameters of the TOI-2195 stellar binary in mind. While future work will study population-level rates in more detail (see Weldon et al. in prep.), the rates we obtain can be taken as suggestive, given the relatively ``average'' location and mass of the stellar perturber. Indeed, a hot Jupiter fraction of $\sim10\%$ was found in \cite{Weldon+26} for a varied stellar population, similar to the result obtained here.} We observe a relative abundance of planets with masses $<0.1M_J$ compared to planets in the range $0.1<m_p<0.3 M_J$, corresponding to the mass at which planets undergoing mass loss tend to stabilize in the structure models (see Appendix \ref{app:numericalmethods}). 
Weldon et al. in prep., explore how the mass distribution of surviving hot Neptunes depends on the specific choice of interior structure model.

In Figure \ref{fig:toi2195pop}, planets consistent with TOI-2195 A b require initial values of $e_c \gtrsim0.5$ and $50^{\circ} \lesssim i \lesssim 130^{\circ}$. Highly eccentric stellar binaries are common, given the uniform distribution that has been observed \citep[e.g.,][]{Raghavan+10,Moe+17}, and a lower $e_c$ may be allowed if the true semimajor axis of the companion is lower than the projected separation, which is likely given that an eccentric object is more probably observed near apoapse. These constraints on the companion's orbit are not surprising given that the EKL mechanism is stronger for systems with a lower ratio of the inner to outer semi-major axes and for more eccentric and inclined systems, as the strength of the octupole contribution to EKL is given by \citep[e.g.,][]{Lithwick+11,Katz+11,Naoz16}
\begin{equation}
    \epsilon_{\rm oct} = \frac{a_p}{a_c} \frac{e_p}{1-e_p^2}  \, .
\end{equation}
A particularly strong EKL effect is necessary to induce significant tidal mass loss to produce a sub-Jovian remnant \citep[][]{Weldon+26}. As demonstrated in the simulations, this scenario is entirely plausible given the known location of the stellar binary companion.

\begin{figure*}
    \centering
    \includegraphics[width=14cm]{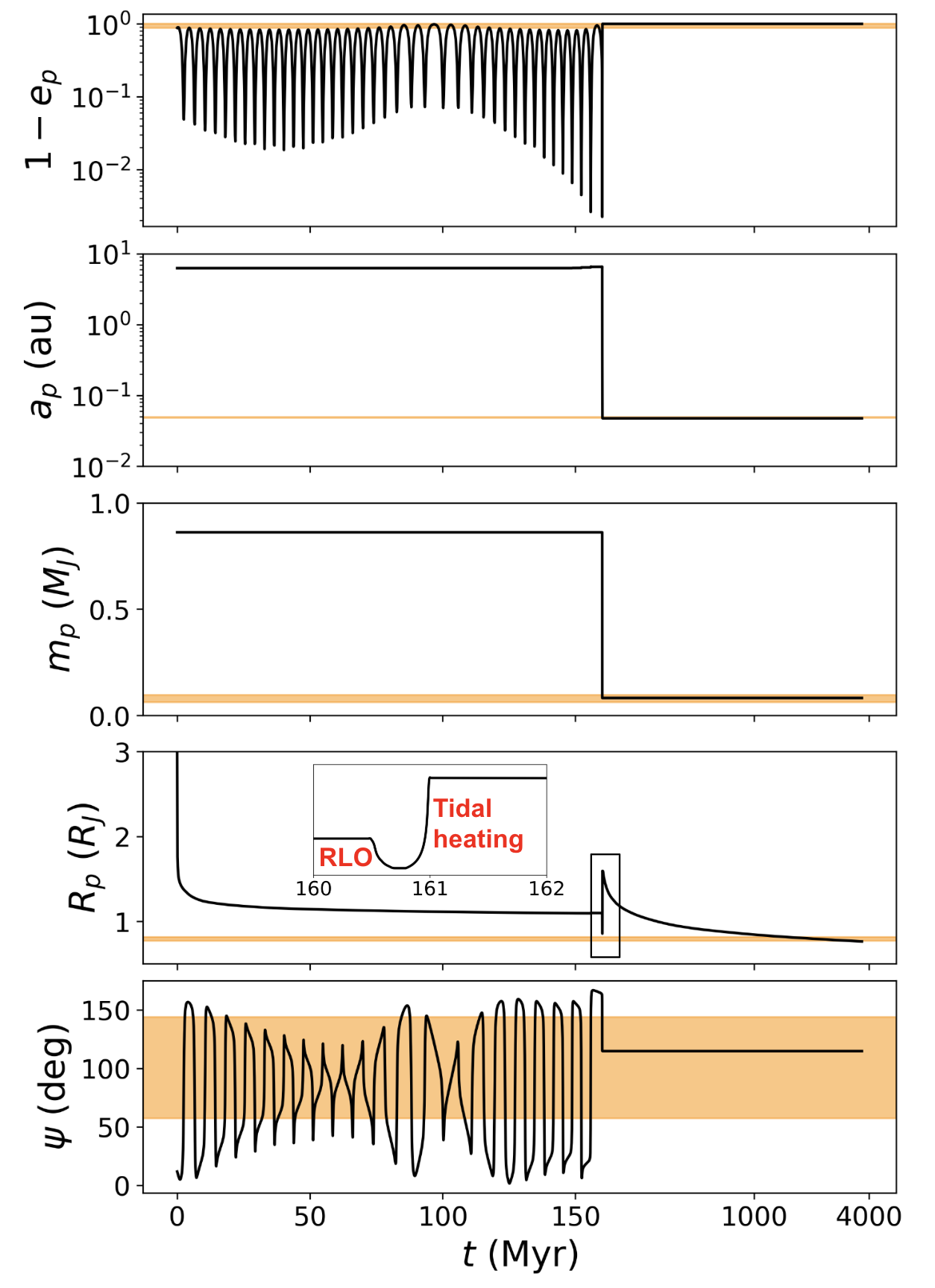}
    \caption{Simulated formation history of a TOI-2195 A b analog. From top to bottom, we show planetary eccentricity ($1-e_p$), semimajor axis ($a_p$), mass ($m_p$), radius ($R_p$), and stellar obliquity ($\psi$). Black curves show the simulated evolution, and orange shaded regions show observational constraints (where $\lambda$ is shown for the constraint on obliquity). The zoomed-in region in the $R_p$ panel shows the radius evolution when the planet loses mass and undergoes radius inflation due to tidal heating. The system initially has $M_* = 0.9 M_{\odot}$, $m_p = 0.862 M_J$, $m_c = 0.5M_{\odot}$, $a_p = 6.27$ au, $a_c = 600$ au, $e_p = 0.119$, $e_c = 0.765$, $\psi = 11.59^{\circ}$, $i=75.61^{\circ}$, $g_1 = 156.88^{\circ}$, and $g_2 = 287.40^{\circ}$.}
    \label{fig:toi2195formation}
\end{figure*}

In Figure \ref{fig:toi2195formation}, we show the time evolution of a system that forms a planet consistent with TOI-2195 A b. The planet begins as a distantly orbiting cold Jupiter ($a_p \sim 6$ au, $m_p \sim 0.9 M_J$) on a low eccentricity, nearly aligned orbit with its host star. The stellar companion induces quadrupolar EKL cycles of eccentricity that gradually increase due to the octupole contribution. At high eccentricities, the planet undergoes Roche lobe overflow to become a stripped sub-Jovian ($m_p\sim0.08M_J)$, and the radius decreases in accordance with the MESA models\footnote{We note that some planets may alternatively inflate during Roche lobe overflow, depending on their mass and entropy in the MESA grid.}. The orbit then undergoes rapid tidal shrinking to $a_p\sim0.05$ au ($t_{\rm shrink } \sim 10^6 \, \rm{yr}$), while the radius inflates due to tidal heating at low semi-major axes ($L_{\rm tide} \propto a_p^{-6}$). The increased radius allows for even more efficient tidal circularization ($L_{\rm tide} \propto R_p^5$), ultimately damping the eccentricity and quenching the inflation of the planet. Planetary cooling and structural response prevent the planet from being altogether disrupted at this stage, although it is in principle possible for some systems. Throughout the migration process, the stellar obliquity chaotically oscillates before "freezing out" at a near-polar value.\footnote{The observed likely polar orbit supports the hypothesis that the planet was excited by EKL, but EKL with mass loss is able to produce a broad distribution of obliquities \citep[][]{Weldon+26}. Therefore, if the true obliquity is lower than that estimated in this work, formation from this channel remains possible.}

During the dynamical evolution, we cool the planet using models appropriate to planets receiving low levels of irradiation, described in Appendix \ref{app:numericalmethods}. Following the migration of the planet, the planet may undergo slower cooling on Gyr timescales due to the high levels of irradiation. Post-circularization, we fix the orbit and model the radius using the publicly available Planetary Structure ANd Dynamics (PSAND) code \citep[][]{Hallatt+25b}, allowing us to accurately assess the radius following the planetary migration and inflation. We match the radius at the end of our dynamical evolution for a planet with a 10$M_{\oplus}$ core and the measured equilibrium temperature of 1098 K reported in this work. We find that the planet's observed radius of $\sim0.079R_J$ is attained after a few Gyr in the PSAND model. Given the degeneracy between age and interior structure and the lack of precise age constraints on the system, our goal is not a precise measurement of the planet's interior structure. Rather, we show that the observed radius is consistent with a highly inflated planet that has cooled on $\sim$Gyr timescales, for a planet with a core+envelope structure consistent with original formation from core accretion \citep[e.g.,][]{Pollack+96}.

After formation, photoevaporation may lead the planet to lose mass \citep[e.g,][]{Owen+13}. We do not include this effect, as our goal is primarily to understand the dynamical formation of the planet. It is possible that the planet ended migration as a slightly more massive Neptune (a scenario allowed by our numerical simulations), before being whittled down to its present-day mass by photoevaporation.

\subsection{Relationship between TOI-2195 A b and the hot Neptune population}

TOI-2195 A b is one of a growing number of hot Neptunes that have been discovered. Here, we discuss the relationship between TOI-2195 A b and other similar systems.

One of the most well-studied analogs to TOI-2195 A b is WASP-107 b, which orbits in the ridge ($P\sim5.7$ days) and has a slightly larger radius ($R_p = 0.94M_J$) and mass ($m_p = 0.12M_J$) \citep[][]{Anderson+17}. While TOI-2195 A b was likely driven to migration by the observed stellar companion, \cite{Yu+24} find that the giant planet companion WASP-107 c could have driven WASP-107 b to undergo high-eccentricity migration. Another well-studied Neptune ridge planet is HAT-P-11 b \citep[][]{Bakos+10}, which has a similar mass to TOI-2195 A b ($\Mp = 0.08\,\Mjup$), but a lower radius ($\Rp = 0.42\,\Rjup$). This planet may have also been driven to high-eccentricity migration by a giant planet companion \citep[][]{Lu+25}. The different radii of these planets may arise from observing the planets at different stages in their cooling history, following tidal heating during migration. More precise age constraints on a large sample of hot Neptune host stars may provide a test for this hypothesis.

TOI-2195 A b likely occupies a highly misaligned orbit, much like several other hot Neptunes that have been discovered \citep[e.g.,][]{Yee+25c,Bourrier+25,Espinoza-Retamal+26}. The stellar obliquity distribution of hot Neptunes resembles that of hot Jupiters, consisting of an aligned population and a population with nearly isotropic orientations \citep[][]{Espinoza-Retamal+26}. \cite{Espinoza-Retamal+26} find that the aligned
component for hot Neptunes appears broader
than that observed for Jupiters below the
Kraft break, suggesting that tidal realignment is more efficient for the more massive Jovians. The correlations between the obliquity distributions suggest that hot Neptunes and hot Jupiters may share a common evolutionary pathway. We propose that high-eccentricity migration with planetary mass loss may account for both populations.

Recent work by \cite{Zanazzi+26} studied the formation of the hot Jupiter pileup and hot Neptune desert and ridge from high-eccentricity migration with dynamical tides. In their model, planets rapidly migrate without significant mass loss to the hot Jupiter pileup and Neptune ridge, whereas in this work the Neptune ridge planets are remnants of partial disruption. Neptune desert planets in \cite{Zanazzi+26} are formed from shocked planets in which the envelope is nearly completely unbound. We do not predict the formation of most Neptune desert planets from Jovian mass loss in this model, as mass-losing planets generally end their evolution at larger orbital periods. An additional distinction is that the model of \cite{Zanazzi+26} does not require the tidal heating and radius inflation of the planet due to tidal circularization, which we show is necessary to produce the puffy radius of TOI-2195 A b and similar planets.

Future constraints on the presence of dynamical tides in the giant planet population, as well as searches for signs of mass loss, such as infrared transients \citep[e.g.,][]{De+23}, may help to disentangle our model for the formation of the Neptune ridge planets from that of \cite{Zanazzi+26}. A parallel observational clue will come from obliquity measurements of Neptune desert dwellers. An aligned obliquity distribution would suggest that these planets form from late timescale Roche lobe overflow (i.e., not during high-eccentricity migration) \citep[][]{Hallatt+26}, whereas a broad obliquity distribution would be indicative of dynamical tidal shocks during high-eccentricity migration \citep[][]{Zanazzi+26}. Observational constraints on the desert population may in turn hint at the origin of more distant planets in the Neptune ridge.

\section{Conclusion} \label{sec:conclusion}

We have presented the discovery of TOI-2195 A b, a Neptune-like planet ($m_p =1.5M_{\rm Nep}$) with a Jupiter-like radius ($R_p = 0.8R_J$) that orbits in the Neptune ridge ($P = 4.16$ days). We also report the 2.6$\sigma$ detection of the Rossiter-McLaughlin effect, indicating that the planet is likely on a near-polar orbit ($\lambda = 109^{+35}_{-53}{^{\circ}}$). A stellar companion is found with a projected separation of $\sim600$ au. This benchmark planet provides a key test for theories of planet formation and migration. We propose that such planets form from the partial disruption of more massive Jovians during high-eccentricity migration.

We perform numerical simulations that solve for the orbital dynamics of the three-body system, coupled with planetary structure evolution, including tidal mass loss and tidal heating. We find that initial Neptunes with fixed mass generally take long timescales to migrate ($>5$ Gyr, as $t_{\rm shrink} \propto m_p/R_p^5$ for fixed $Q_p$), unless they circularize to very short orbital periods ($P \lesssim 3$ days). Therefore, migration as an original Neptune generally cannot account for TOI-2195 A b or other planets in the Neptune ridge ($3<P<6$ days). However, our simulations show that planets in the Neptune ridge are consistent with starting as Jupiters that undergo mass loss to become hot Neptunes. As seen in Figure \ref{fig:migrations}, the Neptune ridge is bounded by the analytic threshold for the tidal disruption of Jovian planets. The resulting high radius and low mass enable rapid migration to the ridge, during which tidal radius inflation occurs. The large observed radius of the planet is consistent with a highly inflated planet that has slowly cooled on Gyr timescales.

TOI-2195 A b is one of a growing population of hot Neptunes. Similarities between the obliquity distribution, metallicity distribution, period distribution, and stellar companionship fraction of these planets with the hot Jupiter population suggests that they share a common evolutionary history \citep[][]{Dong+18,Espinoza-Retamal+26,Vissapragada+25,CastroGonzalez+24,EelesNolle+25}. Here, we show that hot Neptunes may be the remnants of giant planets that undergo tidal mass loss during high-eccentricity migration.

\begin{acknowledgments}
S.W.Y. gratefully acknowledges support from the Heising-Simons Foundation. G.W. and S.N. thank Howard and Astrid Preston for their generous support. G.W. thanks the generous support of the Thacher Fellowship at UCLA. G.W., B.H., and S.N. acknowledge the support of NASA XRP grant 80NSSC23K0262. D.R. was supported by NASA under award number 80NSSC25M7110.

G.W. acknowledges the use of the UCLA cluster \textit{Hoffman2} for computational resources. G.W., B.H., and S.N. thank Expanse at the San Diego Supercomputer Center through ACCESS Discover project PHY250211 \citep[][]{Expanse} for computational resources.

This paper includes data collected by the \TESS mission that are publicly available from the Mikulski Archive for Space Telescopes (MAST).
Funding for the \TESS mission is provided by NASA's Science Mission Directorate. We acknowledge the use of public \TESS data from pipelines at the \TESS Science Office and at the \TESS Science Processing Operations Center.
This research has made use of the Exoplanet Follow-up Observation Program (ExoFOP; DOI: 10.26134/ExoFOP5) website, which is operated by the California Institute of Technology, under contract with the National Aeronautics and Space Administration under the Exoplanet Exploration Program \citep{ExoFoP,ExoFoPTESS}. Resources supporting this work were provided by the NASA High-End Computing (HEC) Program through the NASA Advanced Supercomputing (NAS) Division at Ames Research Center for the production of the SPOC data products.

This research has used data from the CTIO/SMARTS 1.5m telescope, which is operated as part of the SMARTS Consortium by RECONS (\url{www.recons.org}) members Todd Henry, Hodari James, Wei-Chun Jao, and Leonardo Paredes. At the telescope, observations were carried out by Roberto Aviles and Rodrigo Hinojosa.
The CHIRON data were obtained from telescope time allocated under the NN-EXPLORE program with support from the National Aeronautics and Space Administration.

\end{acknowledgments}

\facilities{
TESS, CTIO:1.5m (CHIRON), Magellan:Clay (PFS), LCOGT, Gemini:South (Zorro), SOAR.
}

\bibliography{paperbib,softwarebib,instruments,catalogs,software}{}
\bibliographystyle{aasjournalv7}

\appendix

\section{Numerical simulations}
\label{app:numericalmethods}

We follow a similar procedure to \cite{Weldon+26} to solve for the coupled planetary interior and dynamical evolution. The key difference here is the extension of planetary structure models from 0.3$M_J$ to 0.05$M_J$ to study sub-Jovian planets. We also include the effect of tidal heating, which inflates the planet during late times in the migration.

\subsection{Secular evolution}

We numerically solve the octupole-level secular equations for the hierarchical three-body system following \citet{Naoz+11,Naoz+13}. We also include general relativistic precession of the inner and outer orbit \citep[][]{Naoz+13b}. Equilibrium tides are included following the formalism of \cite{Eggleton98} and \cite{Eggleton+01}. We fix the viscous timescale $t_{v,*} = 50$ years for the star and $t_{v,p} = 0.05$ years for the planet \citep[consistent with intermediate values in e.g.,][]{Petrovich15b}. The stellar and planetary spins are initialized to the solar (24.47 days) and Jovian (0.41 days) values, respectively. The code follows the precession of the spin vector \citep[see for full set of equations,][]{Naoz16}. Furthermore, we model the stellar evolution using the SSE evolution code of \cite{Hurley2000}. The combined code with secular evolution, general relativity, tides, and stellar evolution has been tested and applied to various astrophysical systems in the literature \citep[e.g.,][]{Naoz16,Naoz+16,Stephan+16,Stephan+18,Stephan+19,Stephan+20,Stephan21,Angelo+22,Shariat+23,Shariat+24,Weldon+25,Holzknecht+25,Weldon+26}.

The upper limit for each system's integration time in our simulations of initial Jovians is 10 Gyr. We stop the simulation if a planet is engulfed during stellar evolution, or if the planet's mass falls below $0.05 M_J$ during mass loss. If a hot planet forms, the orbit circularizes until the integration is stopped once $e_p < 10^{-4}$.

All of our initial conditions are stable according to the \cite{Mardling+01} stability criterion
\begin{equation}
    \frac{a_c}{a_p}>2.8\left(1+\frac{m_c}{M_*+m_p}\right)^{2 / 5} \frac{\left(1+e_c\right)^{2 / 5}}{\left(1-e_c\right)^{6 / 5}}\left(1-\frac{0.3 i}{180^{\circ}}\right) \ .
\end{equation}
The systems are also sufficiently hierarchical for stability. That is, the hierarchy parameter
\begin{equation}
\label{eq:eps}
    \epsilon = \frac{a_p}{a_c} \frac{e_c}{1-e_c^2} \ ,
\end{equation}
fulfills $\epsilon<0.1$ \citep[e.g.,][]{Naoz16}.

\subsection{Mass loss and angular momentum return}

We model mass loss at high eccentricities using the procedure of \cite{Weldon+26}. Hydrodynamical simulations by \cite{Guillochon+11} show that planets begin losing mass when the periastron $q$ crosses $2.7r_t$, where the Roche limit $r_t$ is 
\begin{equation} \label{eq:roche}
    r_t = R_p \left(\frac{M_*+m_p}{m_p}\right)^{1/3} \ .
\end{equation}Recent work on the Roche limit modified by dynamical tides found a similar threshold where mass loss may begin \citep[][]{Yu+25}, and more recent hydrodynamical simulations show that planets fractionally lose mass with decreasing $q$ \citep[][]{Fan+26}. As in \cite{Weldon+26}, we fit an exponential to the fractional mass loss in \cite{Guillochon+11} and obtain
\begin{equation}
    \frac{\delta m_p}{m_p} = A \exp\left(-B \frac{q}{r_t}\right)  \ ,
\end{equation}
where $A = 8 \times 10^6$ and $B = 11.5$ \citep[using][]{WebPlotDigitizer}. \cite{Yu+24} used a similar procedure to study the properties of the planet WASP-107 b. To ensure a numerically smooth transition from the regime of no mass loss, in the code, we take 
\begin{equation}
\label{eq:massloss}
    \frac{\delta m_p}{m_p} = A \exp\left(-B \frac{q}{r_t}\right) \left[ 1+\exp \left(250 \left(\frac{q}{r_t}-2.7 \right) \right)\right]\ .  
\end{equation}
Planets may pass from $q/r_t>2.8$ (no mass loss occurs) to $2.7<q/r_t<2.8$ (a negligible, but smooth amount of loss occurs), to $q/r_t<2.7$ (the mass loss follows the exponential fit). 

Mass loss during each individual periapse passage makes secular orbit averaging difficult. Many episodes of mass loss may occur within a large secular time step. Therefore, when a planet falls below $q/r_t = 2.8$ and mass loss begins, we set the integration time step equal to the orbital period of the planet. Following an episode of mass loss, assumed to be at periapse, we treat the specific angular momentum $h$ as conserved to recalculate the planetary orbit \citep[see e.g.,][for similar treatment of mass loss in hierarchical triple systems]{Lu+19}. This approach allows for modeling the planetary and orbital response to mass loss with an orbit-by-orbit resolution.

For each parcel of lost mass $\delta{m_p}$ lost during an individual orbit, a fraction $f_{\rm ret}$ is retained by the system and may be accreted onto the star via torques between the tidal streamers and planet ($\delta{m_p} \ll M_*$). The fraction $f_{\rm ret}$ of angular momentum is returned to the planetary orbit by taking
\begin{equation}
    \frac{dh}{dt} = f_{\rm ret} \frac{\delta{m_p}}{m_p} \frac{r}{r_0} \frac{h_0}{\tau}  \ ,
\end{equation}
where we set $\tau = 100$~yr to be a characteristic timescale for the angular momentum to return (corresponding roughly to the viscous timescale of the accretion flow), $r = a_p(1+e_p^2/2)$ is the time-averaged lever arm for the tangential torque applied on an eccentric orbit, assuming the torque is applied over many orbits. $r_0$ is the lever arm at the time of mass loss, and $h_0$ is the specific angular momentum at the time of mass loss. Over multiple orbits, the torque from many parcels of lost mass accumulates. We denote the cumulative lost mass that a planet experiences over many parcels as $\Delta m_p = \sum \delta m_p$. We stop applying a parcel's torque after the angular momentum of that parcel $f_{\rm ret} \delta m_p h_0$ is completely given back to the orbit.

The fraction $f_{\rm ret}$ of angular momentum returned to the planetary orbit is an observationally ill-constrained quantity; for a systematic exploration of varying $f_{\rm fet}$ from 0 to 1 on the hot Jupiter population, see \cite{Weldon+26}. In this work, we take $f_{\rm ret}=0.15$. This value is found by \cite{Valsecchi+15} and \cite{Hallatt+25} to reconstruct observed features in the population of sub-Jovians formed by late-timescale Roche lobe overflow on nearly circular orbits. It is unclear whether $f_{\rm ret}$ varies with eccentricity, though we may expect the impulsive nature of the mass loss to lead to a low $f_{\rm ret}$. We choose to use the same value of $f_{\rm ret}$ as these studies, as a larger fraction may destabilize the planets that undergo significant mass loss and survive as sub-Jovians in our simulations.  In future work, we will explore systematically how changing $f_{\rm ret}$ affects the production of sub-Jovians.

\subsection{Planetary structure modeling}

\subsubsection{Grid construction}

To account for the planetary response to mass loss, we use Modules for Experiments in Stellar Astrophysics (MESA)
\citep[e.g.,][]{Paxton2011, Paxton2013, Paxton2015, Paxton2018, Paxton2019}, which has the ability to model the evolution of giant planets using the \texttt{make\_planets} module. While it is possible to evolve MESA and the secular code simultaneously \citep[e.g.,][]{Gao+25}, such simulations are computationally prohibitive for large population synthesis studies, especially when accounting for mass loss on an orbit-by-orbit basis. Therefore, a pre-calculated grid allows us to approximate the structural evolution for many systems over a large parameter space. 

We construct MESA models for planets with masses $0.3 M_J < m_p < 10 M_J$, and analytically extend these models into the sub-Jovian regime where \texttt{make\_planets} has difficulty converging, while ensuring physical quantities are consistent at the transition. The planets in this work have a $10 M_{\oplus}$ core, corresponding to the minimum mass for runaway gas accretion \citep[e.g.,][]{Pollack+96}. The remainder of the planet has a solar composition envelope. In MESA, we vary the total planet mass, using masses of 0.3-0.9$M_J$ (spaced in increments of 0.1$M_J$) and 1.0-10.0$M_J$ (spaced in increments of 0.5$M_J$). The initial planet radius set in \texttt{make\_planets} is 2.0$R_J$. To access the mass-radius relations at various entropies, we vary the stellar irradiation levels, considering fluxes of $10^{5}$, $10^{6}$, $10^{7}$, and $5 \times 10^{7}$ erg s$^{-1}$ cm$^{-2}$. We also vary the starting radius of the planet in the $5 \times 10^{7}$ erg s$^{-1}$ cm$^{-2}$ run from 2.0-6.0$R_J$ (in increments of 1.0$R_J$) to access higher entropies.

The MESA \texttt{make\_planets} has difficulty converging below $\sim0.3M_J$, so we analytically solve for the planetary structure in this regime. A small number of models that fail to converge above 0.3$M_J$ are interpolated. For each fixed entropy track ($S=6.5-12.5 $ $k_B/$baryon in increments of 0.5), we extend the MESA mass–radius relation to lower masses using a two-component planetary structure model consisting of a 10$M_{\oplus}$ rocky core and a convective polytropic envelope. The core radius is informed from empirical constraints \citep[e.g.,][]{Chen+17}
\begin{equation}
    R_{\rm core} = 0.97 \left( \frac{m_{\rm core}}{M_{\oplus}} \right)^{0.28} R_{\oplus} \ .
\end{equation}
The envelope is approximated with a polytropic equation of state
\begin{equation}
    P = K \rho^{1+1/n} \, ,
\end{equation}
where both the polytropic index $n$ and constant $K$ are calibrated for each track. The envelope structure is found by integrating hydrostatic equilibrium outward from the core boundary
\begin{equation}
    \frac{dP}{dr} = \frac{-G m(r) \rho}{r^2} 
\end{equation}
\begin{equation}
    \frac{dm}{dr} = 4 \pi r^2 \rho \ .
\end{equation}
The integration starts at $r = R_{\rm core}$, $m = m_{\rm core}$, and $\rho = \rho_b$, and proceeds outward until the pressure reaches a prescribed surface pressure of $10^3$ dyne cm$^{-2}$ (1 mbar), correspondingly roughly to the pressure at the transit radius for observed planets \citep[e.g.,][]{Hubbard+01}. For a given $n$ and $K$, the boundary density $\rho_b$ is adjusted until the integrated total planet mass equals the desired planet mass. For each stitch, $K$ is chosen so that the planetary radius matches at the boundary. Then $n$ is chosen so that the local logarithmic slope of the polytrope extension matches that of the MESA track
\begin{equation}
   \left. \frac{d \ln R_p}{d \ln m_p} \right|_{\rm polytrope} = \left. \frac{d \ln R_p}{d \ln m_p} \right|_{\rm MESA} \, .
\end{equation}

\begin{figure*}[ht]
\begin{center}

\includegraphics[width=4in]{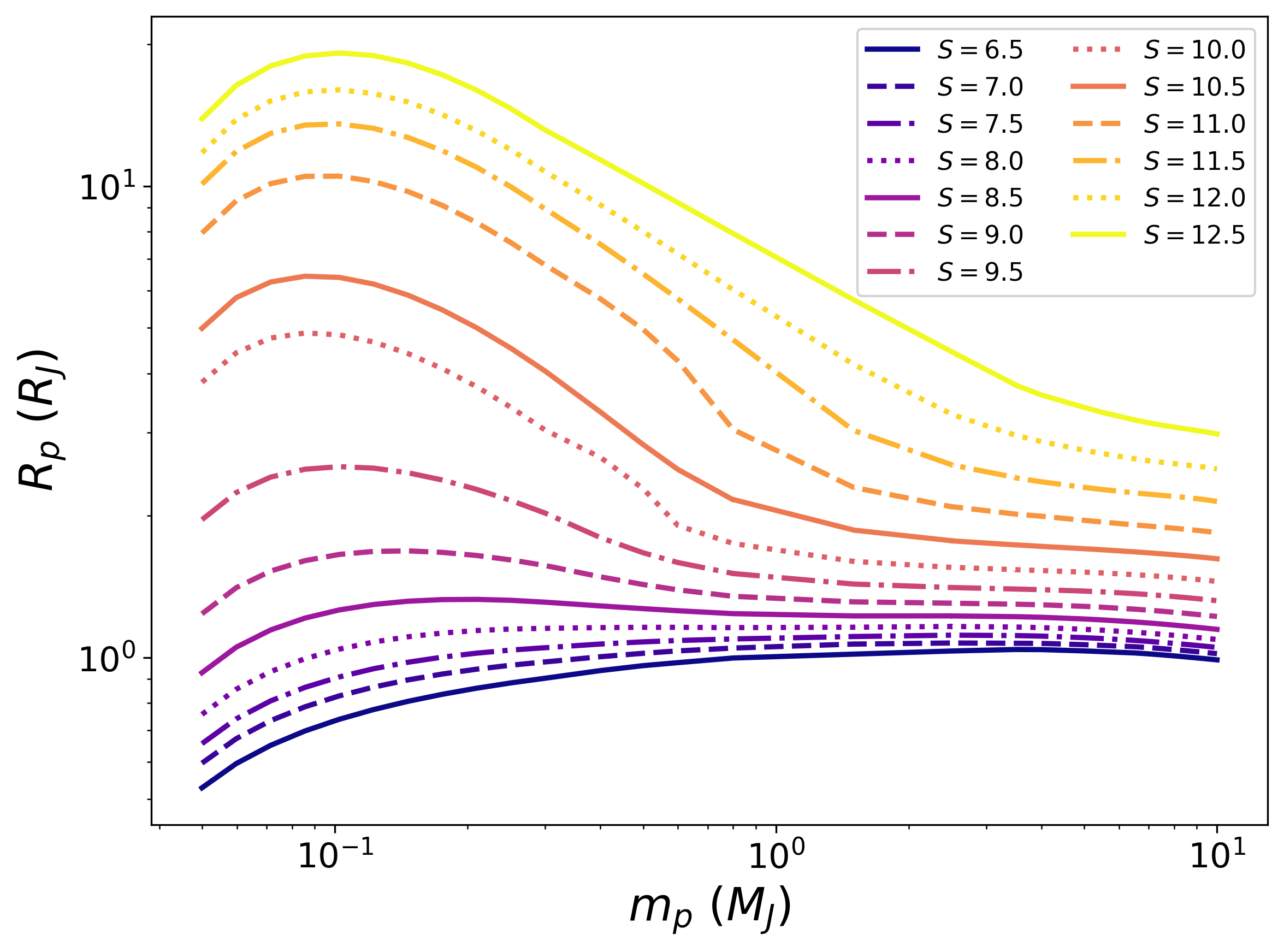}
\caption{\footnotesize Curves of constant central entropy (given in $k_B$/baryon) showing planetary radius as a function of planetary mass. These curves are calculated using MESA in the Jovian regime and by analytically solving for the planetary structure and stitching to the MESA models for the sub-Jovian regime.}
\label{fig:mesa_grids}
\end{center}
\end{figure*}

In Figure \ref{fig:mesa_grids}, we show the planetary structure grid. At the lowest masses and highest entropies, we provide the relations for completeness; however, we expect planets to undergo either runaway mass loss or tidal eccentricity damping before reaching the most extreme inflated radii.

\subsubsection{Stepping through the adiabats}

We use the "stepping through the adiabats" method by evolving the central entropy $S$ of the planet over time \citep[e.g.,][]{Hubbard77,Arras+06,Hallatt+25b}. We then use the planet's mass and entropy to find its radius from the structure grid. As in \cite{Weldon+26}, we treat the entropy as constant during episodes of mass loss to map the new mass to radius, as the mass loss occurs on timescales much shorter than the timescale over which the planet thermally readjusts. We start each planet with a "hot start" entropy of 11 $k_b$/baryon \citep[e.g.,][]{Marleau+14}. When the planet is not actively losing mass, we account for tidal heating and planetary cooling
\begin{equation}
    \label{eq:dSdt}
    \frac{dS}{dt} = \frac{\mu m_{u}}{k_B} \frac{ L_{\rm tide} -L_{\rm cool}}{\int Tdm_p},
\end{equation}
where $k_B$ is the Boltzmann constant, $m_u$ is the atomic mass unit, and $\mu=2.3$ is the approximate mean molecular weight commonly used for giant planets, though this value can vary with composition and structure \citep[e.g.,][]{Pevcnik+05}. The integral $\int Tdm$ is taken over the convective zone of the planet, approximated as the entire envelope in the analytical extension and from the core to the outermost edge of the outer convective zone in the MESA models. It is an open question where tidal heating occurs in giant planets. We apply the heating uniformly over the convective zone, an approach consistent with the idea that dissipation is driven by turbulent viscosity in the bulk fluid \citep[e.g.,][]{Eggleton98}.  

Over the structure grid, we fit for $\int T dm_p$ to obtain this value as a function of mass and entropy
\begin{equation}
    \log_{10} \int Tdm_p = 34.624 + 1.728 \log_{10}(m_p/M_J) + 0.0345 (\log_{10} (m_p/M_J))^2 + 0.0593(S-10) + 0.048 \log_{10}(m_p/M_J) (S-10) \ ,
\end{equation}
for $\int T dm_p$ in units of K g. For a Jupiter-like planet ($m_p = 1M_J$, $S = 6.5$ $k_B$/baryon), this fit gives $\int Tdm \sim 2 \times 10^{34}$ K g, consistent as expected with the product of the mass of Jupiter ($\sim 2 \times 10^{30}$ g) and a mean interior temperature of $\sim 10^4$ K.

The tidal luminosity from eccentricity tides is given by \citep[][]{Leconte+10}
\begin{equation}
    L_{\rm tide} = 2K \left[N_a(e) - \frac{N^2(e)}{\Omega(e)} \right] \, ,
\end{equation}
where 
\begin{equation}
    N_{a}(e) = \frac{1 + \dfrac{31}{2}e^{2} + \dfrac{255}{8}e^{4} + \dfrac{185}{16}e^{6} + \dfrac{25}{64}e^{8}}{(1 - e^{2})^{\frac{15}{2}}} \, ,
\end{equation}
\begin{equation}
    N(e) = \frac{1 + \dfrac{15}{2}e^{2} + \dfrac{45}{8}e^{4} + \dfrac{5}{16}e^{6}}{(1 - e^{2})^{6}} \, ,
\end{equation}
\begin{equation}
    \Omega(e) = \frac{1 + 3e^{2} + \dfrac{3}{8}e^{4}}{(1 - e^{2})^{\frac{9}{2}}} \, ,
\end{equation}
are functions of the eccentricity. The characteristic tidal luminosity scale is 
\begin{equation}
    K = \frac{3n_p}{2} \frac{k_{2,p}}{Q_p} \left(\frac{G M_*^2}{R_p} \right) \left(\frac{R_p}{a_p}\right)^6 \, ,
\end{equation}
where $n_p$ is the mean motion. $k_{2,p}$ and $Q_p$ are parameters relating to the planetary interior structure. We solve for the Love number $k_{2,p}$ from the structure models \citep[following e.g.,][]{Sterne39,Batygin09,Hallatt+25b}:
\begin{equation}
    k_{2,p} = \frac{3-\eta}{2+\eta} \, ,
\end{equation}
where $\eta$ for the entire planet is obtained by integrating $\eta(r)$ radially outward
\begin{equation}
    r \frac{d \eta}{dr} + \eta^2  - \eta - 6 + 6 \frac{\rho}{\rho_m}(\eta+1) = 0 \, ,
\end{equation}
where $\rho_m$ is the mean density interior to $r$.

Over the entire grid, we fit for $k_2$ as a function of mass and entropy to obtain
\begin{equation}
    k_{2,p} (m_p, S) = k_{\rm min} + \frac{k_{\rm max}-k_{\rm min}}{1+ \exp{\frac{S-S_0}{dS}}} \, ,
\end{equation}
where $k_{\rm min} = 0.032$, $k_{\rm max} = 0.542-0.079 \log_{10}{(m_p/M_J)}$, $S_0 = 9.448 + 3.509 \log_{10}{(m_p/M_J)}$, and $dS = 0.865$. For a Jupiter-like planet ($m_p = 1M_J$, $S = 6.5$ $k_B$/baryon), this fit gives $k_{2,p} \sim 0.53$, which is close to the value of $k_{2.p} \sim 0.565$ measured by Juno \citep[e.g.,][]{Durante+20}. Though there are many uncertainties inherent to understanding planetary interior structures, this fit allows us to approximate the direction and magnitude of the tidal response of the planet as its structure changes \citep[see for similar treatment e.g.,][]{Hallatt+25}. Given $k_{2,p}$, we calculate the tidal quality factor $Q_p$ as \citep[e.g.,][]{Fabrycky+07,Hansen10}
\begin{equation}
    Q_p = \frac{4}{3} \frac{k_{2,p}}{2(1+k_{2,p})^2} \frac{G m_p}{R_p^3} \frac{t_{v,p}}{n_p} \, .
\end{equation}
 We manually verify that the integrated tidal luminosity in the simulations is equivalent within a factor of order unity to the change in orbital energy of the planet during its evolution.

We calculate the planetary cooling luminosity using the fits of \cite{Marleau+14}, which they find to be in strong agreement with MESA models. We verify that these fits agree closely with the MESA models that we calculate:
\begin{equation}
    L_{\rm cool, low \, S} = 1.5 \times 10^{-7} L_{\odot} \left( \frac{m_p}{M_J} \right)^{0.72} 10^{1.3(S-8.2)} \ , 
\end{equation}
\begin{equation}
    L_{\rm cool, intermediate \, S} = 7.2 \times 10^{-6} L_{\odot} \left( \frac{m_p}{M_J} \right)^{0.98} 10^{1.58(S-9.6)} \ , 
\end{equation}
\begin{equation}
    L_{\rm cool, high \, S} = 8.7 \times 10^{-5} L_{\odot} \left( \frac{m_p}{M_J} \right)^{0.29} 10^{0.58(S-10.2)} \ , 
\end{equation}
for the low ($S\leq 9$), intermediate ($9<S<10.2$), and high ($S \geq 10.2$) entropy regimes. These fits are well-defined for the Jovian regime, where migrating planets spend the vast majority of time. For sub-Jovian planets on highly irradiated orbits post-migration, we use the publicly available Planetary Structure ANd Dynamics (PSAND) code \citep[][]{Hallatt+25b} to model the late-timescale cooling.

In Eq.~\ref{eq:dSdt}, we neglect to include a stellar irradiation term, as the irradiation generally penetrates shallowly into the photosphere and does not significantly raise the central entropy \citep[e.g.,][]{Wu+13,Komacek+17,Komacek+20}. Indeed, prior studies that have investigated the anomalously large radii of some hot Jupiters found that increased surface heating cannot be the dominant contributor \citep[see for review, e.g.,][]{Thorngren+24}. When the hot Neptune forms and the planet is highly irradiated, we use the PSAND code \citep[][]{Hallatt+25b}, described in Section \ref{sec:formation}.

\section{TESS and High Angular Resolution Imaging Observations of TOI-2195}
\label{app:imaging}

\begin{figure}
    \centering
    \includegraphics[width=0.5\linewidth]{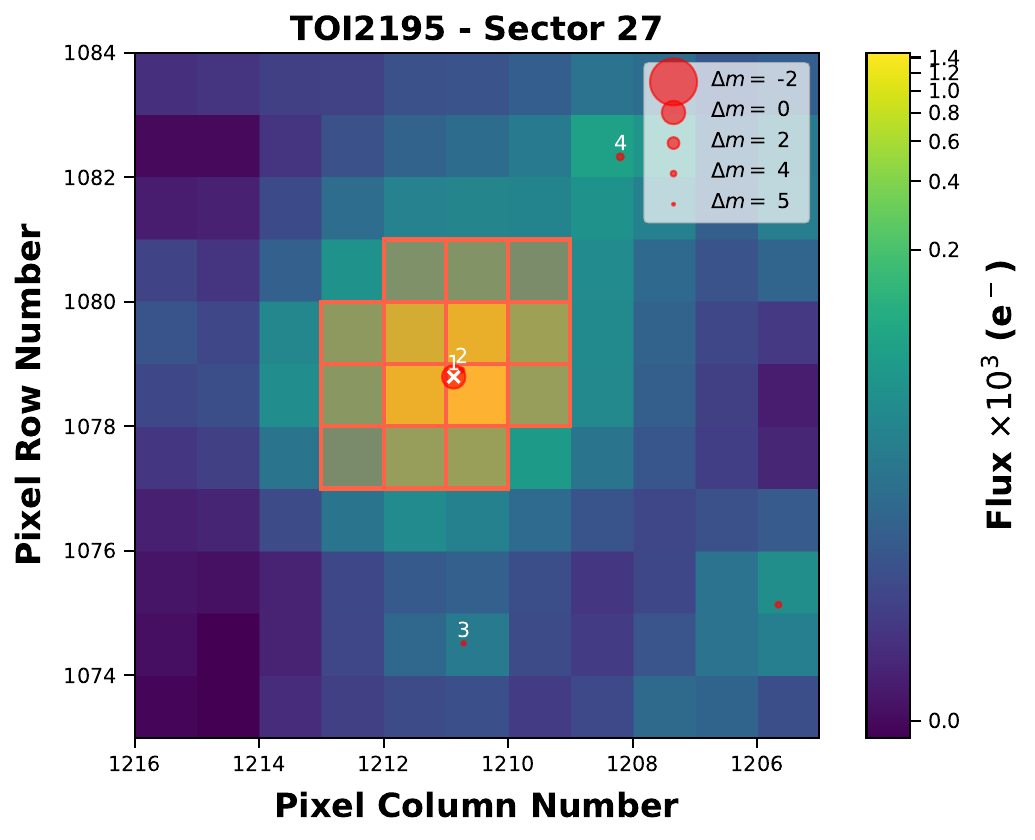}
    \caption{TESS field around TOI-2195, with photometric aperture highlighted in orange. Known stars are shown as red circles, with sizes scaled according to their $G$-band magnitudes relative to TOI-2195. This figure was generated using the publicly available Python package \texttt{tpfplotter} \citep{tpfplotter_Aller2020}. The pixel scale of TESS is 21$^{\prime\prime}$.}
    \label{fig:toi2195_tpf}
\end{figure}

\begin{figure}
    \centering
    \includegraphics[width=0.48\linewidth]{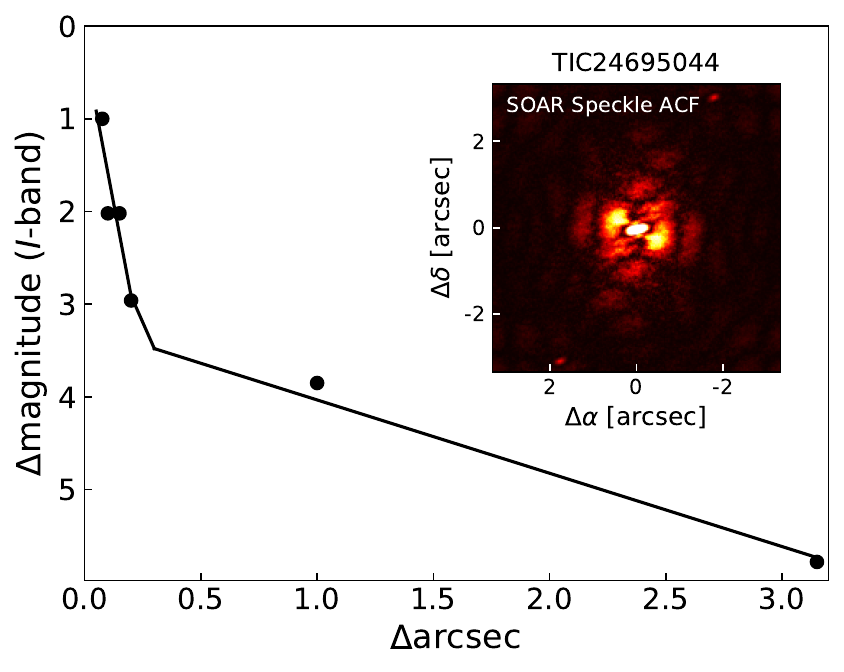}
    \includegraphics[width=0.48\linewidth]{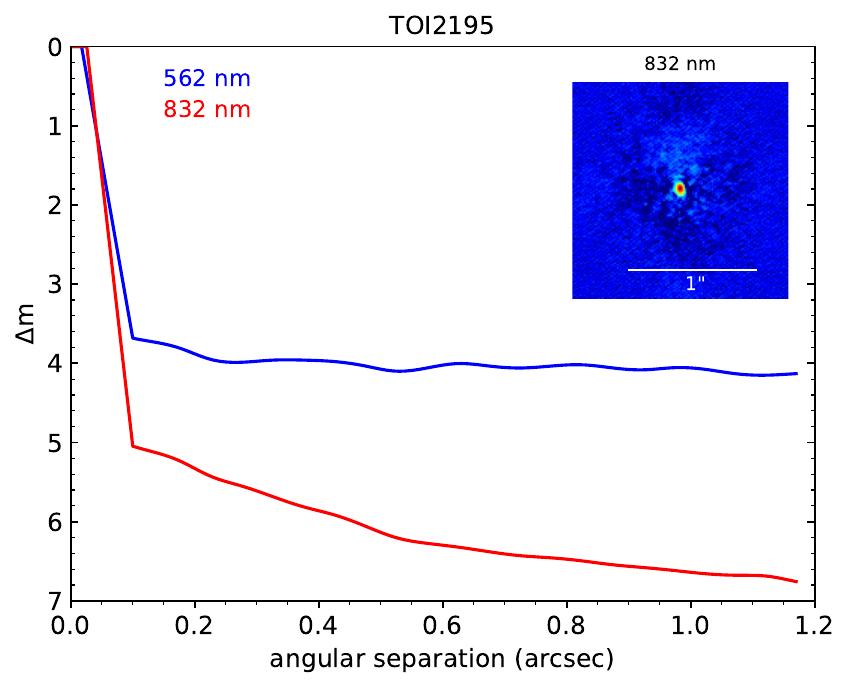}
    \caption{Speckle imaging observations of TOI-2195 from SOAR HRCam (left) and Gemini-South/Zorro (right). Lines show
the 5-sigma magnitude detection limits of each observation and the 832 nm speckle reconstructed image is shown in the inset panel on the right.}
    \label{fig:toi2195_imaging}
\end{figure}

Figure \ref{fig:toi2195_tpf} shows the TESS field of view around TOI-2195 and the photometric aperture used to generate the light
curves. No significant contamination from stars apart from the TOI-2195 binary system is expected within the aperture.
Figure \ref{fig:toi2195_imaging} shows high resolution optical imaging of TOI-2195. TOI-2195 has a fainter companion star at 3 arcsec separation (PA=210 degrees) but to within the angular and magnitude contrast limits of the high-resolution speckle observations, no additional close companion to TOI-2195 was detected.

\end{document}